\documentclass[aps,prl,twocolumn,showpacs,preprintnumbers]{revtex4}
\usepackage{amsmath}
\usepackage{graphicx}
\usepackage{dcolumn}
\usepackage{bm}
\usepackage{subfigure}
\usepackage{amsfonts}
\usepackage{amssymb}

\begin{document}

\title{\textbf{Lattice Boltzmann inverse kinetic approach for the
incompressible Navier-Stokes equations}}
\author{ Enrico Fonda$^{1},$Massimo Tessarotto\thanks{%
Electronic-mail: M.Tessarotto@cmfd.univ.trieste.it}$^{1,2}$ and Marco Ellero$%
^{3}$} \affiliation{$^{1}$Dipartimento di Matematica e
Informatica, Universit\`{a} di Trieste,
Italy\\
$^{2}$Consorzio di Magnetofluidodinamica, Trieste, Italy\\
$^{3}$Institute of Aerodynamics, Technical University of Munich,
Munich,
Germany\\
}
\date{\today }

\begin{abstract}
In spite of the large number of papers appeared in the past which are
devoted to the lattice Boltzmann (LB) methods, basic aspects of the theory
still remain unchallenged. An unsolved theoretical issue is related to the
construction of a discrete kinetic theory which yields \textit{exactly} the
fluid equations, i.e., is non-asymptotic (here denoted as \textit{LB inverse
kinetic theory}). The purpose of this paper is theoretical and aims at
developing an inverse kinetic approach of this type. In principle infinite
solutions exist to this problem but the freedom can be exploited in order to
meet important requirements. In particular, the discrete kinetic theory can
be defined so that it yields exactly the fluid equation also for arbitrary
non-equilibrium (but suitably smooth)\ kinetic distribution functions and
arbitrarily close to the boundary of the fluid domain. This includes the
specification of the kinetic initial and boundary conditions which are
consistent with the initial and boundary conditions prescribed for the fluid
fields. Other basic features are the arbitrariness of the "equilibrium"
distribution function and the condition of positivity\textit{\ }imposed on
the kinetic distribution function.{\small \ }The latter can be achieved by
imposing a suitable \textit{entropic principle,} realized by means of a
constant H-theorem. \ Unlike previous entropic LB methods the theorem can be
obtained without functional constraints on the class of the initial
distribution functions. As a basic consequence, the choice of the the
entropy functional remains essentially arbitrary so that it can be
identified with the Gibbs-Shannon entropy. Remarkably, this property is not
affected by the particular choice of the kinetic equilibrium (to be assumed
in all cases strictly positive). \ Hence, it applies also in the case of
polynomial equilibria, usually adopted in customary LB approaches. We
provide different possible realizations of the theory and asymptotic
approximations which permit to determine the fluid equations \textit{with
prescribed accuracy.} As a result, asymptotic accuracy estimates of
customary LB approaches and comparisons with the Chorin artificial
compressibility method are discussed.
\end{abstract}

 \maketitle

%\bmulticol

\section{1 - Introduction - Inverse kinetic theories}

Basic issues concerning the foundations classical hydrodynamics still remain
unanswered. A remarkable aspect is related the construction of inverse
kinetic theories (IKT) for hydrodynamic equations in which the fluid fields
are identified with suitable moments of an appropriate kinetic probability
distribution. \ The topic has been the subject of theoretical investigations
both regarding the incompressible Navier-Stokes (NS) equations (INSE) \cite%
{Ellero2000,Ellero2004,Ellero2005,Tessarotto2006,Tessarotto2006b,Tessarotto2007}
and the quantum hydrodynamic equations associated to the Schr\"{o}dinger
equation\ \cite{Piero}.\ The importance of the IKT-approach for classical
hydrodynamics goes beyond the academic interest. In fact, INSE represent a
mixture of hyperbolic and elliptic pde's, which are extremely hard to study
both analytically and numerically. As such, their investigation represents a
challenge both for mathematical analysis and for computational fluid
dynamics. The discovery of IKT \cite{Ellero2000} provides, however, a new
starting point for the theoretical and numerical investigation of INSE. In
fact, an inverse kinetic theory yields, by definition, \emph{an exact solver
for the fluid equations}: all the fluid fields, including the fluid pressure
$p(\mathbf{r},t),$ are uniquely prescribed in terms of suitable momenta of
the kinetic distribution function, solution of the kinetic equation. In the
case of INSE this permits, in principle, to determine the evolution of the
fluid fields without solving explicitly the Navier-Stokes equation, nor the
Poisson equations for the fluid pressure \cite{Tessarotto2007}.{\Large \ }%
Previous IKT approaches \cite%
{Ellero2004,Ellero2005,Tessarotto2006,Tessarotto2006b,Piero} have been based
on continuous phase-space models. However, the interesting question arises
whether similar concepts can be adopted also to the development of discrete
inverse kinetic theories based on the lattice Boltzmann (LB) theory. The
goal of this investigation is to propose a novel LB theory for INSE, based
on the development of an IKT with discrete velocities, here denoted as \emph{%
lattice Boltzmann inverse kinetic theory (LB-IKT).} In this paper we intend
to analyze the theoretical foundations and basic properties of the new
approach useful to display its relationship with previous CFD and lattice
Boltzmann methods (LBM) for incompressible isothermal fluids. In particular,
we wish to prove that it delivers an inverse kinetic theory, i.e., that it
realizes an exact Navier-Stokes and Poisson solver.

\subsection{1a - Motivations: difficulties with LBM's}

Despite the significant number of theoretical and numerical papers appeared
in the literature in the last few years, the lattice Boltzmann method \cite%
{McNamara1988,Higueras1989,Succi1991,Benzi1992,ChenpChen-1991,Chen1992,Succi}
- among many others available in CFD\ - is probably the one for which a
complete understanding is not yet available. Although originated as an
extension of the lattice gas automaton \cite{Frisch1986,Frisch1987} or a
special discrete form of the Boltzmann equation \cite{He1997}, several
aspects regarding the very foundation of LB theory still remain to be
clarified. Consequently, also the comparisons and exact relationship between
the various lattice Boltzmann methods (LBM) and other CFD methods are made
difficult or, at least, not yet well understood. Needless to say, these
comparisons are essential to assess the relative value (based on the
characteristic computational complexity, accuracy and stability) of LBM and
other CFD methods. In particular the relative performance of the numerical
methods depend strongly on the characteristic spatial and time
discretization scales, i.e., the minimal spatial and time scale lengths
required by each numerical method to achieve a prescribed accuracy. On the
other hand, most of the existing knowledge of the LBM's properties
originates from numerical benchmarks (see for example \cite%
{Martinez1994,Hou1995,He1997b}). Although these studies have demonstrated
the LBM's accuracy in simulating fluid flows, few comparisons are available
on the relative computational efficiency of the LBM and other CFD methods
\cite{He1997,He2002}. \ The main reason [of these difficulties] is probably
because current LBM's, rather than being exact Navier-Stokes solvers, are at
most asymptotic ones (\emph{asymptotic LBM's}), i.e., \ they depend on one
or more infinitesimal parameters and recover INSE only in an approximate
asymptotic sense.

The motivations of this work are related to some of the basic features of
customary LB theory representing, at the same time, assets and weaknesses.
One of the main reasons of the popularity of the LB approach lays in its
simplicity and in the fact that it provides an approximate Poisson solver,
i.e., it permits to advance in time the fluid fields without explicitly
solving numerically the Poisson equation for the fluid pressure. However
customary LB approaches can yield, at most, only asymptotic approximations
for the fluid fields. This is because of two different reasons. The first
one is the difficulty in the precise definition of the kinetic boundary
conditions in customary LBM's, since sufficiently close to the boundary the
form of the distribution function prescribed by the boundary conditions is
not generally consistent with hydrodynamic equations. The second reason is
that the kinetic description adopted implies either the introduction of weak
compressibility \cite%
{McNamara1988,Higueras1989,Succi,Benzi1992,ChenpChen-1991,Chen1992} or
temperature \cite{Ansumali2002} effects of the fluid or some sort of state
equation for the fluid pressure \cite{Shi2006}. These assumptions, although
physically plausible, appear unacceptable from the mathematical viewpoint
since they represent a breaking of the exact fluid equations.

\ Moreover, in the case of very small fluid viscosity customary LBM's may
become inefficient as a consequence of the low-order approximations usually
adopted and the possible presence of the numerical instabilities mentioned
above. These accuracy limitations at low viscosities can usually be overcome
only by imposing severe grid refinements and strong reductions of the size
of the time step. \ This has the inevitable consequence of raising
significantly the level of computational complexity in customary LBM's
(potentially much higher than that of so-called direct solution methods),
which makes them inefficient or even potentially unsuitable for large-scale
simulations in fluids.

A fundamental issue is, therefore, related to the construction of more
accurate, or higher-order, LBM's{\small ,} applicable for arbitrary values
of the relevant physical (and asymptotic) parameters.{\LARGE \ }However, the
route which should permit to determine them is still uncertain, since the
very existence of an underlying exact (and non-asymptotic) discrete kinetic
theory, analogous to the continuous inverse kinetic theory \cite%
{Ellero2004,Ellero2005}, is not yet known. According to some authors \cite%
{Shan1998,Ansumali12002,Chikatamarla2006} this should be linked to the
discretization of the Boltzmann equation, or to the possible introduction of
weakly compressible and thermal flow models. However, the first approach is
not only extremely hard to implement \cite{Bardow}, since it is based on the
adoption of higher-order Gauss-Hermite quadratures (linked to the
discretization of the Boltzmann equation), but its truncations yield at most
asymptotic theories. Other approaches, which are based on 'ad hoc'
modifications of the fluid equations (for example, introducing
compressibility and/or temperature effects \cite{Ansumali2005}), by
definition cannot provide exact Navier-Stokes solvers.\ \

Another critical issue is related to the numerical stability of LBM's \cite%
{Succi2002},{\small \ }usually attributed to the violation of the condition
of strict positivity (\emph{realizability condition}) for the kinetic
distribution function \cite{Boghosian2001,Succi2002}. Therefore, according
to this viewpoint, a stability criterion should be achieved by imposing the
existence of an H-theorem (for a review see \cite{McCracken2005}). In an
effort to improve the efficiency of LBM numerical implementations and to
cure these instabilities, there has been recently a renewed interest in the
LB theory. Several approaches have been proposed. The first one involves the
adoption of entropic LBM's (ELBM \cite%
{Karlin1998,Karlin1998aa0,Karlin1999,Boghosian2001} in which the equilibrium
distribution satisfies also a maximum principle, defined with respect to a
suitably defined entropy functional. However, usually these methods lead to
non-polynomial equilibrium distribution functions which potentially result
in higher computational complexity \cite{Yong2003} and less numerical
accuracy\cite{Dellar2002}. Other approaches rely on the adoption of multiple
relaxation times \cite{Lallemand2000,Lallemand2003}. However the efficiency,
of these methods is still in doubt. Therefore{\small ,} the search for new
[LB] models, overcoming these limitations{\small ,} remains an important
unsolved task.

\subsection{1b - Goals of the investigation}

The aim of this work is the development of an inverse kinetic theory for the
incompressible Navier-Stokes equations (INSE) which, besides realizing an
exact Navier-Stokes (and Poisson) solver, overcomes {\tiny \ }some of the
limitations of previous LBM's. Unlike Refs. \cite{Ellero2004,Ellero2005},
where a continuous IKT was considered, here we construct a discrete theory
based on the LB velocity-space discretization. \ In such a type of approach,
the kinetic description is realized by a finite number of discrete
distribution functions $f_{i}(\mathbf{r},t)$, for $i=0,k,$ each associated
to a prescribed discrete constant velocity $\boldsymbol{a}_{i}$ and defined
everywhere in the existence domain of the fluid fields (the open set $\Omega
\times I$ )$.$ The configuration space $\Omega $ is a bounded subset of the
Euclidean space $%
%TCIMACRO{\U{211d} }%
%BeginExpansion
\mathbb{R}
%EndExpansion
^{3}$and the time interval $I$ is a subset of $%
%TCIMACRO{\U{211d} }%
%BeginExpansion
\mathbb{R}
%EndExpansion
.$ \ The kinetic theory is obtained as\emph{\ }in \cite%
{Ellero2004,Ellero2005} by introducing an \emph{inverse kinetic equation
(LB-IKE)} which advances in time the distribution function and by properly
defining a correspondence principle, relating a set of velocity momenta with
the relevant fluid fields.

To achieve an IKT for INSE, however, also a proper treatment of the initial
and boundary conditions, to be satisfied by the kinetic distribution
function, must be included. \ In both cases, it is proven that they can be
defined to be \emph{exactly} \emph{consistent} - at the same time - both
with the hydrodynamic equations (which must hold also arbitrarily close to
the boundary of the fluid domain) and with the prescription of the initial
and Dirichlet boundary conditions set for the fluid fields.{\small \ }\
Remarkably, \ both the choice of the initial and equilibrium kinetic
distribution functions and their functional class remain essentially
arbitrary. In other words, provided suitable minimal smoothness conditions
are met by the kinetic distributions function, \ \emph{for arbitrary initial
and boundary kinetic distribution functions, }the relevant{\small \ }moment
equations of the kinetic equation coincide \emph{identically} with the
relevant fluid equations. \ This includes the possibility of defining a
LB-IKT in which the kinetic distribution function is not necessarily a
Galilean invariant.

This arbitrariness is reflected also in the choice of possible "equilibrium"
distribution functions, which remain essentially free in our theory, and can
be made for example in order to achieve minimal algorithmic complexity. A
possible solution corresponds to assume polynomial- type kinetic equilibria,
as in the traditional asymptotic LBM's. These kinetic equilibria are
well-known to be \emph{non-Galilean invariant} with respect to arbitrary
finite velocity translations. Nevertheless, as discussed in detail in Sec.4,
Subsection 4A, although the adoption of Galilei invariant kinetic
distributions is in possible, this choice does not represent an obstacle for
the formulation of a LB-IKT. Actually Galilean invariance need to be
fulfilled only by the fluid equations. The same invariance property must be
fulfilled only by the moment equations of the LB-IKT and not necessarily by
the whole LB inverse kinetic equation (LB-IKE).

Another significant development of the theory is the formal introduction of
an entropic principle, realized by\ a constant H-theorem, in order to assure
the strict positivity of the kinetic distribution function in the whole
existence domain $\Omega \times I$. \ \ The present entropic principle
departs significantly from the literature. Unlike previous entropic LBM's it
is obtained without imposing any functional constraints on the class of the
initial kinetic distribution functions\textsl{. }Namely without demanding
the validity of a principle of entropy maximization (PEM, \cite{Jaynes1957})
in a true functional sense on the form of the distribution function. Rather,
it follows imposing a constraint only on a suitable set of \textit{extended
fluid fields}, in particular the \textit{kinetic pressure }$p_{1}(\mathbf{r}%
,t)$.The latter is uniquely related to the actual fluid pressure $p(\mathbf{r%
},t)$ via the equation $p_{1}(\mathbf{r},t)=p(\mathbf{r},t)+P_{o}(t),$ with $%
P_{o}(t)>0$ to be denoted as pseudo-pressure. The constant H-theorem is
therefore obtained by suitably prescribing the function $P_{o}(t)$ and
implies the strict positivity. The same prescription assures that the
entropy results maximal with respect in the class of the admissible kinetic
pressures, i.e., it satisfies a principle of entropy maximization. \ \
Remarkably, since this property is not affected by the particular choice of
the kinetic equilibrium, the H-theorem applies also in the case of
polynomial equilibria. \ We stress{\small \ }that the choice of the entropy
functional remains essentially arbitrary, since no actual physical
interpretation can be attached to it. For example, without loss of
generality it can always be identified with the Gibbs-Shannon entropy. Even
prescribing these additional properties, in principle infinite solutions
exist to the problem. Hence, the freedom can be exploited to satisfy further
requirements (for example, mathematical simplicity, minimal algorithmic
complexity, etc.). Different possible realizations of the theory and
comparisons with other CFD approaches are considered. The formulation of the
inverse kinetic theory is also useful in order to determine the precise
relationship between the LBM's and previous CFD schemes and in particular to
obtain possible improved asymptotic LBM's with prescribed accuracy. As an
application, we intend to construct asymptotic models which satisfy with
prescribed accuracy the required fluid equations [INSE] and possibly extend
also the range of validity of traditional LBM's. In particular, this permits
to obtain asymptotic accuracy estimates of customary LB approaches. The
scheme of presentation is as follows. In Sec.2 the INSE problem is recalled
and the definition of the extended fluid fields $\left\{ \mathbf{V}%
,p_{1}\right\} $ is presented. \ In Sec. 3 the basic assumptions of previous
asymptotic LBM's are recalled. \ In.Sec.4 and 5 the foundations of the new
inverse kinetic theory are laid down and the integral LB inverse kinetic
theory is presented, while in Sec. 6 the entropic theorem is proven to hold
for the kinetic distribution function for properly defined kinetic pressure.
Finally, in Sec.7 various asymptotic approximations are obtained for the
inverse kinetic theory and comparisons are introduce with previous LB and
CFD methods and in Sec. 8 the main conclusions are drawn.

\section{2 - The INSE problem}

A prerequisite for the formulation of an inverse kinetic theory \cite%
{Ellero2004,Ellero2005} providing a phase-space description of a classical
(or quantum) fluid is the proper identification of the complete set of fluid
equations and of the related fluid fields. For a Newtonian incompressible
fluid, referred to an arbitrary inertial reference frame, these are provided
by the incompressible Navier-Stokes equations (INSE) for the fluid fields $%
\left\{ \rho ,\mathbf{V,}p\right\} $
\begin{eqnarray}
&&\left. \nabla \cdot \mathbf{V}=0,\right.  \label{INSE-1} \\
&&\left. N\mathbf{V}=\mathbf{0},\right.  \label{INSE-2} \\
&&\rho (\mathbf{r,}t\mathbb{)}=\rho _{o}.  \label{INSE-3}
\end{eqnarray}%
There are supplemented by the inequalities
\begin{eqnarray}
&&\left. p(\mathbf{r,}t)\geq 0,\right.  \label{INSE-4} \\
&&\left. \rho _{o}>0.\right.  \label{INSE-5}
\end{eqnarray}%
Equations (\ref{INSE-1})-(\ref{INSE-3}) are defined in a open connected set $%
\Omega \subseteq \mathbb{R}^{3}$ (defined as the subset of $\mathbb{R}^{3}$
where $\rho (\mathbf{r,}t\mathbb{)}>0$) with boundary $\delta \Omega ,$
while Eqs. (\ref{INSE-4}) and (\ref{INSE-5}) apply on its closure $\overline{%
\Omega }.$ Here the notation is standard. Thus, $N$ is the NS operator
\begin{equation}
N\mathbf{V\equiv }\rho _{o}\frac{D}{Dt}\mathbf{V}+\mathbf{\nabla }p+\mathbf{f%
}-\mu \nabla ^{2}\mathbf{V,}  \label{Eq.0}
\end{equation}%
with $\frac{D}{Dt}=\frac{\partial }{\partial t}+\mathbf{V\cdot \nabla }$ the
convective derivative, $\mathbf{f}$ denotes a suitably smooth volume force
density acting on the fluid element and $\mu \equiv \nu \rho _{o}>0$ is the
constant fluid viscosity. In particular we shall assume that $\mathbf{f}$
can be represented in the form%
\begin{equation*}
\mathbf{f}=\mathbf{-\nabla }\Phi (\mathbf{r})+\mathbf{f}_{1}(\mathbf{r,}t)
\end{equation*}%
where we have separated the conservative $\nabla \Phi (r)$\ and the
non-conservative $f_{1}$ parts of the force. Equations (\ref{INSE-1})-(\ref%
{INSE-3}) are assumed to admit a strong solution in $\Omega \times I,$ with $%
I\subset \mathbb{R}$ a possibly bounded time interval. By assumption $%
\left\{ \rho ,\mathbf{V,}p\right\} $ are continuous in the closure $%
\overline{\Omega }.$ Hence if in $\Omega \times I,$ $\mathbf{f}$ is at least
$C^{(1,0)}(\Omega \times I),$ it follows necessarily that $\left\{ \mathbf{V,%
}p\right\} $ must be at least $C^{(2,1)}(\Omega \times I).$ In the sequel we
shall impose on $\left\{ \mathbf{V,}p\right\} $ the initial conditions
\begin{eqnarray}
\mathbf{V}(\mathbf{r,}t_{o}\mathbb{)} &\mathbf{=}&\mathbf{V}_{o}(\mathbf{r}%
\mathbb{)},  \label{INSE-R1} \\
p(\mathbf{r},t_{o}) &=&p_{o}(\mathbf{r}).  \notag
\end{eqnarray}%
Furthermore, for greater mathematical simplicity, here we shall impose
Dirichlet boundary conditions on $\delta \Omega $
\begin{equation}
\left\{
\begin{array}{ccc}
\left. \mathbf{V(\cdot ,}t\mathbb{)}\right\vert _{\delta \Omega } & = &
\left. \mathbf{V}_{W}\mathbf{(\cdot ,}t\mathbb{)}\right\vert _{\delta \Omega
} \\
\left. p\mathbf{(\cdot ,}t\mathbb{)}\right\vert _{\delta \Omega } & = &
\left. p_{W}\mathbf{(\cdot ,}t\mathbb{)}\right\vert _{\delta \Omega }.%
\end{array}%
\right.  \label{INSE-R2}
\end{equation}%
Eqs.(\ref{INSE-3}) and (\ref{INSE-R1})-(\ref{INSE-R2}) define the
initial-boundary value problem associated to the reduced INSE (\emph{reduced
INSE problem}). It is important to stress that the previous problem can also
formulated in an equivalent way by replacing the fluid pressure $p(\mathbf{r}%
,t)$ with a function $p_{1}(\mathbf{r},t)$ (denoted \emph{kinetic pressure})
of the form
\begin{equation}
p_{1}(\mathbf{r},t)=P_{o}+p(\mathbf{r},t),  \label{Eq.6}
\end{equation}%
where $P_{o}=P_{o}(t)$ is prescribed (but arbitrary) real function of time%
{\small \ }and is at least $P_{o}(t)\in C^{(1)}(I).$ $\left\{ \mathbf{V,}%
p_{1}\right\} $ will be denoted hereon as \emph{extended fluid fields }and $%
P_{o}(t)$ will be denoted as \emph{pseudo-pressure}.

\section{3 - Asymptotic LBM's}

\subsection{3A - Basic assumptions}

As is well known, all LB methods are based on a discrete kinetic theory,
using a so-called lattice Boltzmann velocity discretization of phase-space (%
\emph{LB discretization}). This involves the definition of a kinetic
distribution function $f,$ which can only take the values belonging to a
finite discrete set $\left\{ f_{i}(\mathbf{r},t),i=0,k\right\} $ (\emph{%
discrete kinetic distribution functions}). In particular, it is assumed that
the functions $f_{i},$ for $i=0,k,$ are associated to a discrete set of $k+1$
different "velocities" $\left\{ \mathbf{a}_{i},i=0,k\right\} .$ Each $%
\mathbf{a}_{i}$ is an 'a priori' prescribed constant vector spanning the
vector space $%
%TCIMACRO{\U{211d} }%
%BeginExpansion
\mathbb{R}
%EndExpansion
^{n}$ (with $n=2$ or $3$ respectively for the treatment of two- and
three-dimensional fluid dynamics)$,$and each $f_{i}(\mathbf{r},t)$ is
represented by a suitably smooth real function which is defined and
continuous in $\overline{\Omega }\times I$ and in particular is at least $%
C^{(k,j)}(\Omega \times I)$ with $k\geq 3.$

The crucial aspect which characterizes customary LB approaches \cite%
{McNamara1988,Higueras1989,Succi1991,Benzi1992,ChenpChen-1991,Chen1992,Cao1997,He1997,Abe1997,Succi}
involves the construction of kinetic models which allow a finite sound speed
in the fluid and hence are based on the assumption of a (weak)
compressibility of the same fluid. This is realized by assuming that the
evolution equation (kinetic equation) for the discrete distributions $f_{i}(%
\mathbf{r},t)$ ($i=1,k$), \emph{\ depends at least one (or more)
infinitesimal (asymptotic) parameters (see below)}. \ Such approaches are
therefore denoted as asymptotic LBM's. They are characterized by a suitable
set of assumptions, which typically include:

\begin{enumerate}
\item \emph{LB assumption \#1: discrete kinetic equation and correspondence
principle:} the first assumption concerns the definition of an appropriate
evolution equation for each $f_{i}(\mathbf{r},t)$ which must hold (together
with all its moment equations) in the whole open set $\Omega \times I.$\ In
customary LB approaches it takes the form of the so-called \emph{LB-BGK
equation }\cite{Chen1992,Quian1992,Cao1997}%
\begin{equation}
L_{(i)}f_{i}=\Omega _{i}(f_{i}),  \label{Eq.1}
\end{equation}%
where $i=0,k.$ Here $L_{(i)}$ is a suitable streaming operator,
\begin{equation}
\Omega _{i}(f_{i})=-\nu _{c}(f_{i}-f_{i}^{eq})  \label{Eq.2a}
\end{equation}%
(with $\nu _{c}\geq 0$ a constant \emph{collision frequency}) is known as
BKG\ collision operator (after Bhatbagar, Gross and Krook \cite{BGK}) and $%
f_{i}^{eq}$ is an "equilibrium" distribution to be suitably defined. In
customary LBM's it is implicitly assumed that the solution of Eq.(\ref{Eq.1}%
), subject to suitable initial and boundary conditions exists and is unique
in the functional class indicated above. In particular, usually $L_{(i)}$ is
either identified with the \emph{finite difference streaming operator} (see
for example \cite{McNamara1988,ChenpChen-1991,Quian1992,Chen1992}), i.e., $%
L_{(i)}f_{i}(\mathbf{r},t)=L_{FD(i)}f_{i}(\mathbf{r},t)\equiv \frac{1}{%
\Delta t}\left[ f_{i}(\mathbf{r+a}_{i}\Delta t,t+\Delta t)-f_{i}(\mathbf{r}%
,t)\right] $ or with the \emph{differential streaming operator} (see for
instance \cite{Cao1997,He1997,Abe1997})
\begin{equation}
L_{(i)}=L_{D(i)}\equiv \frac{\partial }{\partial t}+\mathbf{a}_{i}\cdot
\frac{\partial }{\partial \mathbf{r}}.  \label{Eq.2b}
\end{equation}%
Here the notation is standard. In particular, in the case of the operator $%
L_{FD(i)}$, $\Delta t$ and $c\Delta t\equiv L_{o}$ are appropriate
parameters which define respectively the characteristic time- and length-
scales associated to the LBM time and spatial discretizations. A common
element to all LBM's is the assumption that all relevant fluid fields can be
identified, at least in some approximate sense, with appropriate momenta of
the discrete kinetic distribution function (\emph{correspondence principle}%
). In particular, for neutral and isothermal incompressible fluids, for
which the fluid fields are provided respectively by the velocity and
pressure fluid fields $\left\{ Y_{j}(\mathbf{r},t),j=1,4\right\} \equiv
\left\{ \mathbf{V}(\mathbf{r},t),p(\mathbf{r},t)\right\} ,$ it is assumed
that they are identified with a suitable set of discrete velocity momenta
(for $j=1,4$)
\begin{equation}
Y_{j}(\mathbf{r},t)=\sum\limits_{i=0,k}X_{ji}(\mathbf{r},t)f_{i}(\mathbf{r}%
,t),  \label{Eq.-2}
\end{equation}%
where $X_{ji}(\mathbf{r},t)$ (with $i=0,k$ and $j=1,k$) are appropriate,
smooth real weight functions. \ \ In the literature several examples of
correspondence principles are provided, a particular case being provided by
the so-called D2Q9 $(\mathbf{V},p)$-scheme \cite{Xiaoyi1996,Zou1997}\
\begin{eqnarray}
&&\left. p(\mathbf{r},t)=c^{2}\sum\limits_{i=0,k}f_{i}=c^{2}\sum%
\limits_{i=0,k}f_{i}^{(eq)}\right. ,  \label{Eq-3} \\
&&\left. \mathbf{V}(\mathbf{\mathbf{r},}t)=\frac{3}{\rho _{o}}%
\sum\limits_{i=1,k}\mathbf{a}_{i}f_{i}=\frac{3}{\rho _{o}}\sum\limits_{i=1,k}%
\mathbf{a}_{i}f_{i}^{(eq)}\right. ,  \label{Eq-4}
\end{eqnarray}%
where $k=8$ and $c=\min \left\{ \left\vert \mathbf{a}_{i}\right\vert
>0,i=0,k\right\} $ is a characteristic parameter of the kinetic model to be
interpreted as test particle velocity.{\Large \ }In customary LBM's the
parameter $c_{s}=\frac{c}{\sqrt{D}}$\ (with $D$\ the dimension of the set $%
\Omega $) is interpreted as sound speed of the fluid.{\Large \ }In order
that the momenta (\ref{Eq-3}) and (\ref{Eq-4}) recover (in some suitable
approximate sense) INSE , however, appropriate subsidiary conditions must be
met.

\item \emph{LB assumption \#2:} \emph{Constraints and asymptotic conditions:
\ }these are based on\ the introduction of a dimensionless parameter $%
\varepsilon $, to be considered infinitesimal, in terms of which all
relevant parameters can be ordered. In particular, it is required that the
following asymptotic orderings \cite{Cao1997,He1997,Abe1997} apply
respectively to the fluid fields $\rho _{o},\mathbf{V}(\mathbf{r},t),p(%
\mathbf{r},t)$, the kinematic viscosity $\nu =\mu /\rho _{o}$ and Reynolds
number $R_{e}=LV/\nu $:%
\begin{eqnarray}
&&\left. \rho _{o},\mathbf{V}(\mathbf{r},t),p(\mathbf{r},t)\sim
o(\varepsilon ^{0}),\right.  \label{2bb-a} \\
&&\left. \nu =\frac{c^{2}}{3\nu _{c}}\left[ 1+o(\varepsilon )\right] \sim
o(\varepsilon ^{\alpha _{R}}),\right.  \label{2bb-b} \\
&&\left. R_{e}\sim 1/o(\varepsilon ^{\alpha _{R}}),\right.  \label{2bb-c}
\end{eqnarray}%
where $\alpha _{R}\geq 0.$ Here we stress that the position for $\nu $ holds
in the case of D2Q9 only, while the generalization to 3D and other LB
discretizations. is straightforward. Furthermore, the velocity $c$ and
collision frequency $\nu _{c}$ are ordered so that
\begin{eqnarray}
&&\left. c\sim 1/o(\varepsilon ^{\alpha _{c}}),\right.  \label{2bb-1} \\
&&\left. \nu _{c}\sim 1/o(\varepsilon ^{\alpha _{\nu }}),\right.
\label{2bb-2} \\
&&\left. \frac{c}{L\nu _{c}}\sim o(\varepsilon ^{\alpha }),\right.
\label{2bb-3}
\end{eqnarray}%
with $\alpha \equiv \alpha _{\nu }-\alpha _{c}>0;$ the characteristic length
and time scales, $L_{o}\equiv c\Delta t$ and $\Delta t$ for the spatial and
time discretization are assumed to scale as%
\begin{eqnarray}
&&\left. \frac{c\Delta t}{L}\equiv \frac{L_{o}}{L}\sim o(\varepsilon
^{\alpha _{L}}),\right.  \label{2bb-4} \\
&&\left. \frac{\Delta t}{T}\sim o(\varepsilon ^{\alpha _{t}}),\right.
\label{2bb-5}
\end{eqnarray}%
with $\alpha _{t},\alpha _{L}>0.$ Here $L$ and $T$ are the (smallest)
characteristic length and time scales, respectively for spatial and time
variations of $\mathbf{V}(\mathbf{r},t)$ and $p(\mathbf{r}.t)$. Imposing
also that $\frac{1}{T\nu _{c}}$ results infinitesimal at least of order
\begin{equation*}
\frac{1}{T\nu _{c}}\sim o(\varepsilon ^{\alpha })
\end{equation*}%
it follows that it must be also $\alpha _{t}-\alpha _{L}>0.$ These
assumptions imply necessarily that the dimensionless parameter $%
M^{eff}\equiv \frac{V}{c}$\ (Mach number) must be ordered as%
\begin{equation}
M^{eff}\sim O(\varepsilon ^{\alpha _{c}})  \label{Eq.3a}
\end{equation}%
(\emph{small Mach-number expansion}).

\item \emph{LB assumption \#3: Chapman-Enskog expansion - Kinetic initial
conditions, relaxation conditions:} it is assumed that the kinetic
distribution function $f_{i}(r,t)$\ admits a convergent Chapman-Enskog
expansion of the form%
\begin{equation}
f_{i}=f_{i}^{eq}+\delta f_{i}^{(1)}+\delta ^{2}f_{i}^{(2)}+..,  \label{Eq.2}
\end{equation}%
where $\delta \equiv \varepsilon ^{\alpha }$\ and the functions $f_{i}^{(j)}$%
\ ($j\in N$) are assumed smooth functions of the form (multi-scale
expansion) $f_{i}^{(j)}(r_{o},r_{1},r_{2},..t_{o},t_{1},t_{2},..),$\ where $%
r_{n}=\delta ^{n}r,t_{n}=\delta ^{n}t$\ and $n\in
%TCIMACRO{\U{2115} }%
%BeginExpansion
\mathbb{N}
%EndExpansion
.$\ In typical LBM's the parameter $\delta $\ is usually identified with $%
\varepsilon $\ (which requires letting $\alpha =1$), while the
Chapman-Enskog expansion is usually required to hold at least up to order $%
o(\delta ^{2})$. In addition the initial conditions
\begin{equation}
f_{i}(\mathbf{r},t_{o})=f_{i}^{eq}(\mathbf{r},t_{o}),
\end{equation}%
(for $i=0,k$) are imposed in the closure of the fluid domain $\overline{%
\Omega }.$ It is well known \cite{Skordos1993} that this position generally
(i.e., for non-stationary fluid fields), implies the violation of the
Chapman-Enskog expansion close to $t=t_{o}$, since the approximate fluid
equations are recovered only letting $\delta f_{i}^{(1)}+\delta
^{2}f_{i}^{(2)}\neq 0,$ i.e., assuming that the kinetic distribution
function has relaxed to the Chapman-Enskog form (\ref{Eq.2}).\ This implies
a numerical error (in the evaluation of the correct fluid fields) which can
be overcome only discarding the first few time steps in the numerical
simulation.

\item \emph{LB assumption \#5:} \emph{Equilibrium kinetic distribution: \ }a
possible realization for\emph{\ }the equilibrium distributions $f_{i}^{eq}$ (%
$i=0,k$) is given by a polynomial of second degree in the fluid velocity
\cite{Xiaoyi1996}%
\begin{eqnarray}
&&\left. f_{i}^{eq}(\mathbf{r},t)=w_{i}\frac{1}{c^{2}}\left[ p-\Phi (\mathbf{%
r})\right] +\right.  \label{Polynomial} \\
&&+w_{i}\rho _{o}\left[ \frac{\mathbf{a}_{i}\cdot \mathbf{V}}{c^{2}}+\frac{3%
}{2}\left( \frac{\mathbf{a}_{i}\cdot \mathbf{V}}{c^{2}}\right) ^{2}-\frac{1}{%
2}\frac{V^{2}}{c^{2}}\right] .  \notag
\end{eqnarray}%
Here, without loss of generality, the case of the D2Q9 LB discretization
will be considered, with $w_{i}$ and $\mathbf{a}_{i}$ (for $i=0,8$) denoting
prescribed dimensionless constant weights and discrete velocities. Notice
that, by definition, $f_{i}^{eq}$ is \emph{not} a Galilei scalar.
Nevertheless, it can be considered approximately invariant, at least with
respect to low-velocity translations which do not violate the low-Mach
number assumption (\ref{Eq.3a}).

\item \emph{LB assumption \#6:} \emph{Kinetic} \emph{boundary conditions: \ }%
They are specified by suitably prescribing the form of the incoming
distribution function at the boundary $\delta \Omega .$ \cite%
{Ziegler1993,Comumbert1993,Ginzbourg1994,Chen1996,Noble1995,Ladd1994,Noble1995,Zou1996,Maier1996,Chen1996,Mei1999,Bouzidi2001,Ansumali2003b,Ginzburg2003,Junk2005}%
. However, this position is not generally consistent with the Chapman-Enskog
solution (\ref{Eq.2}) (see related discussion in Appendix A).\ As a
consequence violations of the hydrodynamic equations may be expected
sufficiently close to the boundary, a fact which may be only alleviated (but
not completely eliminated) by adopting suitable grid refinements near the
boundary. An additional potential difficulty is related to the condition of
strict positivity of the kinetic distribution function \cite{Ansumali2003b}
which is not easily incorporated into the no-slip boundary conditions \cite%
{Ladd1994,Noble1995,Zou1996}.
\end{enumerate}

\subsection{3B - Computational complexity of asymptotic LBM's}

The requirements posed by the validity of these hypotheses may strongly
influence the computational complexity of asymptotic LBM's which is usually
associated to the total number of "logical"\ operations which must be
performed during a prescribed time interval. Therefore, a critical parameter
of numerical simulation methods is their discretization time scale $\Delta
t. $ This is - in turn - related to the Courant number $N_{C}=\frac{V\Delta t%
}{L_{o}},$ where $V$ and $L_{o}$.denote respectively the sup of the
magnitude of the fluid velocity and the amplitudes of the spatial
discretization. \ As is well known "optimal" CFD simulation methods
typically allow $L_{o}\sim L$ and a definition of the time step $\Delta
t=\Delta t_{Opt}$ such that $N_{C}\sim \frac{V\Delta t_{Opt}}{L}\sim 1$.
Instead, for usual LBM's satisfying the low-$M^{eff}$ assumption (\ref{Eq.3a}%
), the Courant number is very small since\ it results $N_{C}=M^{eff}\frac{%
L_{o}}{L}\sim O(\varepsilon ^{\alpha })\frac{L_{o}}{L}$. This means that
their discretization time scale of $\Delta t$ is much smaller than $\Delta
t_{Opt}$ and reads%
\begin{equation}
\Delta t\sim M^{eff}\frac{L_{o}}{L}\Delta t_{Opt}.  \label{Eq.4b}
\end{equation}%
In addition, depending on the accuracy of the numerical algorithms adopted
for the construction of the discrete kinetic distribution function,\ also
the ratio $\frac{L_{o}}{L}$ results infinitesimal in the sense $\frac{L_{o}}{%
L}$ $\sim o(\varepsilon ^{\alpha _{L}}),$ with suitable $\alpha _{L}>0$.
Finally, we stress that LB approaches based on the adoption of the
finite-difference streaming operator $L_{FD(i)}$ are usually only accurate
to order $o(\Delta t^{2}).$ For them, therefore, the requirement placed by
Eq.(\ref{Eq.4b}) might be even stronger. This implies that traditional LBM's
may involve a vastly larger computation time than that afforded by more
efficient numerical methods.

\section{4 - New LB inverse kinetic theory (LB-IKT)}

A basic issue in LB approaches \cite%
{McNamara1988,ChenpChen-1991,Quian1992,Chen1992} concerns the choice of the
functional class of the discrete kinetic distribution functions $f_{i}$ ($%
i=0,k$) as well as the related definition of the equilibrium discrete
distribution function $f_{i}^{eq}$ [which appears in the BGK collision
operator; see Eq.(\ref{Eq.2a})]. This refers in particular to their
transformation properties\ with respect to arbitrary Galilean
transformations, and specifically to their Galilei invariance with respect
to velocity translations with constant velocity.

In statistical mechanics it is well known that the kinetic distribution
function is usually assumed to be a Galilean scalar. The same assumption
can, in principle, be adopted also for LB models. However, the kinetic
distribution functions $f_{i}$ and $f_{i}^{eq}$ do not necessarily require a
physical interpretation of this type. In the sequel we show that for a
discrete inverse kinetic theory it is sufficient that $f_{i}$ and $%
f_{i}^{eq} $ be so defined that the moment equations coincide with the fluid
equations (which by definition are Galilei covariant). It is sufficient to
demand that both $f_{i}$ and $f_{i}^{eq}$ are identified with a ordinary
scalars with respect to the group of rotation in $%
%TCIMACRO{\U{211d} }%
%BeginExpansion
\mathbb{R}
%EndExpansion
^{2},$ while they need not be necessarily invariant with respect to
arbitrary velocity translations. This means that $f_{i}$ is invariant only
for a particular subset of inertial reference frames. For example for a
fluid which at the initial time moves locally with constant velocity an
element of this set can be identified with the inertial frame which in the
same position is locally co-moving with the fluid.

The adoption of non-translationally invariant discrete distributions $f_{i}$
is actually already well known in LBM and results convenient for its
simplicity. This means, manifestly, that in general no obvious physical
interpretation can be attached to the other momenta of the discrete kinetic
distribution function. \ As a consequence, the very definition of the
concept of statistical entropy to be associated to the $f_{i}^{\prime }$s is
essentially arbitrary, as well as the related principle of entropy
maximization, typically used for the determination of the equilibrium
distribution \ function $f_{i}^{eq}.$ Several authors, nevertheless, have
investigated the adoption of possible alternative formulations, which are
based on suitable definitions of the entropy functional and/or the
requirement of approximate or exact Galilei invariance (see for example \cite%
{Karlin1998,Succi2002,Boghosian2003}).

\subsection{4A - Foundations of LB-IKT}

As previously indicated, there are several important motivations for seeking
an exact solver based on LBM. The lack of a theory of this type represents
in fact a weak point of LB theory. Besides being a still unsolved
theoretical issue, the problem is relevant in order to determine the exact
relationship between the LBM's and traditional CFD schemes based on the
direct discretization of the Navier--Stokes equations. Following ideas
recently developed \cite%
{Ellero2004,Ellero2005,Tessarotto2006,Tessarotto2006b,Piero}, we show that
such a theory can be formulated by means of an inverse kinetic theory (IKT)
with discrete velocities. By definition such an IKT should yield \emph{%
exactly} the complete set of fluid equations and which, contrary to
customary kinetic approaches in CFD (in particular LB methods), should not
depend on asymptotic parameters. This implies that the inverse kinetic
theory must also satisfy an \emph{exact closure condition}. As a further
condition, we require that the fluid equations are fulfilled independently
of the initial conditions for the kinetic distribution function (to be
properly set) and should hold for arbitrary fluid fields. The latter
requirement is necessary since we must expect that the validity of the
inverse kinetic theory should not be limited to a subset of possible fluid
motions nor depend on special assumptions, like a prescribed range of
Reynolds numbers. In principle a phase-space theory, yielding an inverse
kinetic theory, may be conveniently set in terms of a quasi-probability,
denoted as kinetic distribution function, $f(\mathbf{x},t).$ A particular
case of interest (investigated in Refs.\cite{Ellero2004,Ellero2005}) refers
to the case in which $f(\mathbf{x},t)$ can actually be identified with a
phase-space probability density. In the sequel we address both cases,
showing that, to a certain extent, in both cases the formulation of a
generic IKT can actually be treated in a similar fashion. \ This requires
the introduction of an appropriate set of \emph{constitutive assumptions}
(or axioms). These concern in particular the definitions of the kinetic
equation - denoted as \emph{inverse kinetic equation (IKE)} - which advances
in time $f(\mathbf{x},t)$ and of the velocity momenta to be identified with
the relevant fluid fields (\emph{correspondence principle}). However,
further assumptions, such as those involving the regularity conditions for $%
f(\mathbf{x},t)$ and the prescription of its initial and boundary conditions
must clearly be added. The concept [of IKT] can be easily extended to the
case in which the kinetic distribution function takes on only discrete
values in velocity space. In the sequel we consider for definiteness the
case of the so-called \emph{LB discretization}, whereby - for each $\left(
\mathbf{r},t\right) \in $ $\Omega \times I$ \ - the kinetic distribution
function is discrete, and in particular admits a finite set of discrete
values $f_{i}(\mathbf{r},t)\in \mathbb{R},$ for $i=0,k,$ each one
corresponding to a prescribed constant discrete velocity $\mathbf{a}_{i}\in
\mathbb{R}^{3}$ for $i=0,k$.

\subsection{4B - Constitutive assumptions}

Let us now introduce the constitutive assumptions (\emph{axioms}) set for
the construction of a LB-IKT for INSE, whose form is suggested by the
analogous continuous inverse kinetic theory \cite{Ellero2004,Ellero2005}.
The axioms, define the "generic" form of the discrete kinetic equation, its
functional setting, the momenta of the kinetic distribution function and
their initial and boundary conditions, are the following ones:

\subsubsection{\emph{Axiom I} \emph{-} \emph{LB--IKE and functional setting.}%
}

Let us require that the extended fluid fields $\left\{ \mathbf{V,}%
p_{1}\right\} $ are strong solutions of INSE, with initial and boundary
conditions (\ref{INSE-R1})-(\ref{INSE-R2}) and that the pseudo pressure $%
p_{o}(t)$ is an arbitrary, suitably smooth, real function. In particular we
impose that the fluid fields and the volume force belong to the \emph{%
minimal functional setting}:

\begin{eqnarray}
&&p_{1},\Phi \epsilon C^{(2,1)}(\Omega \times I),  \notag \\
&&\mathbf{V}\epsilon C^{(3,1)}(\Omega \times I),  \label{Eq.6a} \\
&&\mathbf{f}_{1}\epsilon C^{(1,0)}(\Omega \times I).  \notag
\end{eqnarray}

We assume that in the set $\Omega \times I$ the following equation
\begin{equation}
L_{D(i)}f_{i}=\Omega _{i}(f_{i})+S_{i}  \label{Eq.7}
\end{equation}%
[\emph{LB inverse kinetic equation (LB-IKE)}]\emph{\ }is satisfied
identically by the discrete kinetic distributions $f_{i}(\mathbf{r},t)$ for $%
i=0,k.$ Here $\Omega _{i}(f_{i})$ and $L_{D(i)}$ are respectively the BGK
and the differential streaming and operators [Eqs.(\ref{Eq.2a}) and (\ref%
{Eq.2b})], while $S_{i}$ is a source term to be defined.\ We require that
KB-IKE is defined in the set $\Omega \times I,$ so that $\Omega _{i}(f_{i})$
and $S_{i}$ are at least that $C^{(1)}(\Omega \times I)\ $and continuous in $%
\overline{\Omega }\times I.$ Moreover $\Omega _{i}(f_{i})$, defined by Eq.(%
\ref{Eq.2a}),\ is considered for generality and will be useful for
comparisons with customary LB approaches. \ We remark that the choice of the
equilibrium kinetic distribution $f_{i}^{eq}$ in the BGK operator remains
completely arbitrary. We assume furthermore that in terms of $f_{i}$ the
fluid fields $\left\{ \mathbf{V},p_{1}\right\} $ are determined by means of
functionals of the form $M_{X_{j}}\left[ f_{i}\right] =\sum%
\limits_{i=0,8}X_{j}f_{i}$ (denoted as \emph{discrete velocity momenta})$.$
For $X=X_{1},X_{2}$ (with $X_{1}=c^{2},X_{2}=\frac{3}{\rho _{o}}\mathbf{a}%
_{i}$) these are related to the fluid fields by means of the equations (%
\emph{correspondence principle})%
\begin{eqnarray}
&&\left. p_{1}(\mathbf{r},t)-\Phi (\mathbf{r})=c^{2}\sum%
\limits_{i=0,8}f_{i}=c^{2}\sum\limits_{i=0,8}f_{i}^{eq}\right. ,
\label{Eq.9} \\
&&\left. \mathbf{V}(\mathbf{\mathbf{r},}t)\mathbf{=}\frac{3}{\rho _{o}}%
\sum\limits_{i=1,8}\mathbf{a}_{i}f_{i}=\frac{3}{\rho _{o}}\sum\limits_{i=1,8}%
\mathbf{a}_{i}f_{i}^{eq}\right. ,  \label{Eq.10}
\end{eqnarray}%
where $c=\min \left\{ \left\vert \mathbf{a}_{i}\right\vert ,\text{ }%
i=1,8\right\} $ is the test particle velocity and{\LARGE \ }$f_{i}^{eq}$\ is
defined by Eq.(\ref{Polynomial}) but with the kinetic pressure $p_{1}$\ that
replaces the fluid pressure $p$\ adopted previously \cite{Xiaoyi1996}. These
equations are assumed to hold identically in the set $\overline{\Omega }%
\times I$ and by assumption, $f_{i}$ and $f_{i}^{eq}$ belong to the same
functional class of real functions defined so that the extended fluid fields
belong to the minimal functional setting (\ref{Eq.6a}). Moreover, without
loss of generality, we consider the D2Q9 LB discretization.

\subsubsection{\emph{Axiom II - Kinetic initial and boundary conditions.}}

The discrete kinetic distribution function satisfies, for $i=0,k$ and for
all $\mathbf{r}$ belonging to the closure $\overline{\Omega }$, the initial
conditions%
\begin{equation}
f_{i}(\mathbf{r},t_{o})=f_{oi}(\mathbf{r,}t_{o})  \label{Eq.10a}
\end{equation}%
where $f_{oi}(\mathbf{r,}t_{o})$ (for $i=0,k$) is a initial distribution
function defined in such a way to satisfy in the same set the initial
conditions for the fluid fields
\begin{eqnarray}
&&p_{1o}(\mathbf{r})\equiv P_{o}(t_{o})+p_{o}(\mathbf{r})-\Phi (\mathbf{r})=
\label{Eq.10b} \\
&&\left. =c^{2}\sum\limits_{i=0,8}f_{oi}(\mathbf{r}),\right.  \notag \\
&&\left. \mathbf{V}_{o}\mathbb{(}\mathbf{\mathbf{r}})=\frac{3}{\rho _{o}}%
\sum\limits_{i=1,8}\mathbf{a}_{i}f_{oi}(\mathbf{r})\right. .  \label{Eq.10c}
\end{eqnarray}%
To define the analogous kinetic boundary conditions on $\delta \Omega ,$ let
us assume that $\delta \Omega $ is a smooth, possibly moving, surface. Let
us introduce the velocity of the point of the boundary determined by the
position vector $\mathbf{r}_{w}\in \delta \Omega ,$ defined by $\mathbf{V}%
_{w}(\mathbf{r}_{w}(t),t)=\frac{d}{dt}\mathbf{r}_{w}(t)$ and denote by $%
\mathbf{n}(\mathbf{r}_{w},t)$ the outward normal unit vector, orthogonal to
the boundary $\delta \Omega $ at the point $\mathbf{r}_{w}.$ Let us denote
by $f_{i}^{(+)}(\mathbf{r}_{w},t)$ and $f_{i}^{(-)}(\mathbf{r}_{w},t)$ the
kinetic distributions which carry the discrete velocities $\mathbf{a}_{i}$
for which there results respectively $\left( \mathbf{a}_{i}-\mathbf{V}%
_{w}\right) \cdot \mathbf{n}(\mathbf{r}_{w},t)>0$ (outgoing-velocity
distributions) and $\left( \mathbf{a}_{i}-\mathbf{V}_{w}\right) \cdot
\mathbf{n}(\mathbf{r}_{w},t)\leq 0$ (incoming-velocity distributions) and
which are identically zero otherwise. We assume for definiteness that both
sets, for which $\left\vert \mathbf{a}_{i}\right\vert >0,$ are non empty
(which requires that the parameter $c$ be suitably defined so that $%
c>\left\vert \mathbf{V}_{w}\right\vert $). The boundary conditions are
obtained by prescribing the incoming kinetic distribution $f_{i}^{(-)}(%
\mathbf{r}_{w},t),$ i.e., imposing (for all $\left( \mathbf{r}_{w},t\right)
\in \delta \Omega \times I$)
\begin{equation}
f_{i}^{(-)}(\mathbf{r}_{w},t)=f_{oi}^{(-)}(\mathbf{r}_{w},t).  \label{Eq.11a}
\end{equation}%
Here $f_{oi}^{(-)}(\mathbf{r}_{w},t)$ are suitable functions, to be assumed
non-vanishing and defined only for incoming discrete velocities for which $%
\left( \mathbf{a}_{i}-\mathbf{V}_{w}\right) \cdot \mathbf{n}(\mathbf{r}%
_{w},t)\leq 0$. Manifestly, the functions $f_{oi}^{(-)}(\mathbf{r}_{w},t)$ ($%
i=0,k$) must be defined so that the Dirichlet boundary conditions for the
fluid fields are identically fulfilled, namely there results
\begin{eqnarray}
&&\left. p_{1w}(\mathbf{r}_{w},t)=P_{o}(t)+p_{w}(\mathbf{r}_{w},t)-\Phi (%
\mathbf{r})=\right.  \label{Eq.11b} \\
&&\left. =c^{2}\sum\limits_{i=0,k}\left\{ f_{oi}^{(-)}(\mathbf{r}%
_{w},t)+f_{i}^{(+)}(\mathbf{r}_{w},t)\right\} ,\right.  \notag \\
&&\left. \mathbf{V}_{w}(\mathbf{\mathbf{r}}_{w},t)=\right.  \label{Eq.11c} \\
&&\left. =\frac{3}{\rho _{o}}\sum\limits_{i=1,k}\mathbf{a}_{i}\left\{
f_{oi}^{(-)}(\mathbf{r}_{w},t)+f_{i}^{(+)}(\mathbf{r}_{w},t)\right\} .\right.
\notag
\end{eqnarray}%
Here, again, the functions $f_{oi}(\mathbf{r})$ and $f_{oi}^{(\pm )}(\mathbf{%
r}_{w},t)$ (for $i=0,k$) must be assumed suitably smooth. A particular case
is obtained imposing identically for $i=0,k$
\begin{eqnarray}
f_{oi}(\mathbf{r,}t_{o}) &=&f_{i}^{eq}(\mathbf{r},t_{o}),  \label{Eq.11d} \\
f_{oi}^{(\pm )}(\mathbf{r}_{w},t) &=&f_{i}^{eq}(\mathbf{r}_{w},t),
\label{Eq.11e}
\end{eqnarray}%
where the identification with $f_{oi}^{(+)}(\mathbf{r}_{w},t)$ and $%
f_{oi}^{(-)}(\mathbf{r}_{w},t)$ is intended respectively in the subsets $%
\mathbf{a}_{i}\cdot \mathbf{n}(\mathbf{r}_{w},t)>0$ and $\mathbf{a}_{i}\cdot
\mathbf{n}(\mathbf{r}_{w},t)\leq 0$. Finally, we notice that in case Neumann
boundary conditions are imposed on the fluid pressure, Eq.(\ref{Eq.11b})
still holds provided $p_{w}(\mathbf{r}_{w},t)$ is intended as a calculated
value.

\subsubsection{\emph{Axiom III -} \emph{Moment equations.}}

If $f_{i}(\mathbf{r},t),$ for $i=0,k,$ are arbitrary solutions of LB-IKE
[Eq.(\ref{Eq.7})] which satisfy Axioms I and II validity of Axioms I and II,
we assume that the moment equations of the same LB-IKE, evaluated in terms
of the moment operators $M_{X_{j}}\left[ \cdot \right] =\sum%
\limits_{i=0,8}X_{j}\cdot ,$ with $j=1,2,$ coincide identically with INSE,
namely that there results identically [for all $\left( \mathbf{r},t\right)
\in \Omega \times I$]
\begin{equation}
M_{X_{1}}\left[ L_{i}f_{i}-\Omega _{i}(f_{i})-S_{i}\right] =\nabla \cdot
\mathbf{V}=0,  \label{Eq.11}
\end{equation}%
\begin{equation}
M_{X_{2}}\left[ L_{i}f_{i}-\Omega _{i}(f_{i})-S_{i}\right] =N\mathbf{V}=%
\mathbf{0.}  \label{Eq.12}
\end{equation}

\subsubsection{\emph{Axiom IV - Source term.}}

The source term is required to depend on a finite number of momenta of the
distribution function. It is assumed that these include, at most, the
extended fluid fields $\left\{ \mathbf{V,}p_{1}\right\} $ and the kinetic
tensor pressure%
\begin{equation}
\underline{\underline{\mathbf{\Pi }}}=3\sum_{i=0}^{8}f_{i}\mathbf{a}_{i}%
\mathbf{a}_{i}-\rho _{o}\mathbf{VV}.  \label{Eq.13}
\end{equation}

\begin{itemize}
\item Furthermore,\ we also normally require\emph{\ }(except for the LB-IKT
described in Appendix B)\emph{\ }that $S_{i}(\mathbf{r},t)$ results
independent of $f_{i}^{eq}(\mathbf{r,}t),$ $f_{oi}(\mathbf{r})$ and $f_{wi}(%
\mathbf{r}_{w},t)$ (for $i=0,k$).
\end{itemize}

Although, the implications will made clear in the following sections, it is
manifest that these axioms do not specify uniquely the form (and functional
class) of the equilibrium kinetic distribution function $f_{i}^{eq}(\mathbf{%
r,}t),$ nor of the initial and boundary kinetic distribution functions (\ref%
{Eq.10a}),(\ref{Eq.11a}). Thus, both $f_{i}^{eq}(\mathbf{r,}t),f_{oi}(%
\mathbf{r,}t_{o})$ and the related distribution they still remain in
principle \emph{completely arbitrary}. \ Nevertheless, by construction, the
initial and (Dirichlet) boundary conditions for the fluid fields are
satisfied identically. \ In the sequel we show that these axioms define a
(non-empty) family of parameter-dependent LB-IKT's, depending on two
constant free parameters $\nu _{c},c>0$ and one arbitrary real function $%
P_{o}(t).$ The examples considered are reported respectively in the
following Sec. 5,6 and in the Appendix B.

\section{5 - A possible realization: the integral LB-IKT}

We now show that, for arbitrary choices of the distributions $f_{i}(\mathbf{%
r,}t)$ and $f_{i}^{eq}(\mathbf{r,}t)$ which fulfill axioms I-IV$,$ an
explicit (and non-unique) realization of the LB-IKT can actually be
obtained. We prove, in particular, that a possible realization of the
discrete inverse kinetic theory, to be denoted as \emph{integral LB-IKT, }is
provided by the source term

\begin{eqnarray}
&&\left. S_{i}=\right.  \label{Eq.14} \\
&\equiv &\frac{w_{i}}{c^{2}}\left[ \frac{\partial p_{1}}{\partial t}-\mathbf{%
a}_{i}\cdot \left( \mathbf{f}_{1}\mathbf{-}\mu \mathbf{\nabla }^{2}\mathbf{V}%
-\nabla \cdot \underline{\underline{\mathbf{\Pi }}}+\nabla p\right) \right]
\equiv \widetilde{S}_{i},  \notag
\end{eqnarray}%
where $\frac{w_{i}}{c^{2}}\frac{\partial p_{1}}{\partial t}$ is denoted as
first pressure term. Holds, in fact, the following theorem.

\subsection{Theorem 1 - \emph{Integral LB-IKT}}

\emph{In validity of axioms I-IV the following statements hold. For an
arbitrary particular solution }$f_{i}$\emph{\ and for arbitrary extended
fluid fields}$:$\emph{\ }

\emph{A) if }$\ f_{i}$\ \emph{\ is a solution of LB-IKE [Eq.(\ref{Eq.7})]
the moment equations coincide identically with INSE in the set }$\Omega
\times I;$

\emph{B) the initial conditions and the (Dirichlet) boundary conditions for
the fluid fields are satisfied identically;}

\emph{C) in validity of axiom IV the source term }$\widetilde{S}_{i}$\emph{\
is non-uniquely defined by Eq.(\ref{Eq.14}).}

\textbf{Proof}

\emph{A)} We notice that by definition there results identically

\begin{equation}
\sum_{i=0}^{8}\widetilde{S}_{i}=\frac{1}{c^{2}}\frac{\partial p_{1}}{%
\partial t}  \label{Eq.19}
\end{equation}%
$\mathbf{\qquad }$%
\begin{eqnarray}
&&\left. \sum_{i=0}^{8}\mathbf{a}_{i}\widetilde{S}_{i}=\right.  \label{Eq.20}
\\
&&\left. =-\frac{1}{3}\left[ \mathbf{f-}\mu \mathbf{\nabla }^{2}\mathbf{V-}%
\nabla \cdot \underline{\underline{\mathbf{\Pi }}}+\nabla p\right] \right.
\notag
\end{eqnarray}%
\ On the other hand, by construction (Axiom I) $f_{i}$ ($i=1,k$) is defined
so that there results identically $\sum_{i=0}^{8}\Omega _{i}=0$ and $%
\sum_{i=0}^{8}\mathbf{a}_{i}\Omega _{i}=\mathbf{0}.$ Hence the momenta $%
M_{X_{1}},M_{X_{2}}$ of LB-IKE deliver respectively
\begin{equation}
\nabla \cdot \sum\limits_{i=1,8}\mathbf{a}_{i}f_{i}=0  \label{21}
\end{equation}%
\begin{equation}
3\frac{\partial }{\partial
t}\sum\limits_{i=1,8}\mathbf{a}_{i}f_{i}+\rho
_{o}\mathbf{V\cdot \nabla V+\nabla }p_{1}+\mathbf{f-}\mu \mathbf{\nabla }^{2}%
\mathbf{V}=\mathbf{0}  \label{22}
\end{equation}%
where the fluid fields $\mathbf{V,}p_{1}$ are defined by Eqs.(\ref{Eq.9}),(%
\ref{Eq.10}). Hence Eqs.(\ref{21}) and (\ref{22}) coincide respectively with
the isochoricity and Navier-Stokes equations [(\ref{INSE-1}) and (\ref%
{INSE-2})]. As a consequence, $f_{i}$ is a particular solution of LB-IKE iff
the fluid fields $\left\{ \mathbf{V,}p_{1}\right\} $ are strong solutions of
INSE.

\emph{B)} Initial and boundary conditions for the fluid fields are satisfied
identically by construction thanks to Axiom II.

\emph{C) }However, even prescribing $\nu _{c},c>0$ and the real function $%
P_{o}(t)$, the functional form of the equation cannot be unique The non
uniqueness of the functional form of the source term $\widetilde{S}_{i}(%
\mathbf{r},t)$ is assumed to be independent of $f_{i}^{eq}(\mathbf{r,}t)$
[and hence of Eq.(\ref{Eq.7})] is obvious. In fact, let us assume that $%
\widetilde{S}_{i}$ is a particular solution for the source term which
satisfies the previous axioms I-IV. Then, it is always possible to add to $%
S_{i}$ arbitrary terms of the form $\widetilde{S}_{i}+\delta S_{i},$ with $%
\delta S_{i}\neq 0$ which depends only on the momenta indicated above, and
gives vanishing contributions to the first two moment equations, namely $%
M_{X_{j}}\left[ \delta S_{i}\right] =\sum\limits_{i=0,8}X_{j}\delta S_{i}=0,$
with $j=1,2$. \ To prove the non-uniqueness of the source term $S_{i}$, it
is sufficient to notice that, for example, any term of the form $\delta
S_{i}=\left( \frac{3}{2}\frac{a_{i}^{2}}{c^{2}}-1\right) F(\mathbf{r},t)$,
with $F(\mathbf{r},t)$ an arbitrary real function (to be assumed, thanks to
Axiom IV, a linear function of the fluid velocity), gives vanishing
contributions to the momenta $M_{X_{1}},M_{X_{2}}.$ Hence $\widetilde{S}_{i}$
is non-unique.

The implications of the theorem are straightforward. First, manifestly, it
holds also in the case in which the BGK operator vanishes identically. This
occurs letting $\nu _{c}=0$ in the whole domain $\Omega \times I.$ Hence the
inverse kinetic equation holds independently of the specific definition of $%
\ f_{i}^{eq}(\mathbf{r,}t).$

An interesting feature of the present approach lies in the choice of the
boundary condition adopted for $f_{i}(\mathbf{r,}t),$ which is different
from that usually adopted in LBM's [see for example \cite{Succi} for a
review on the subject]. In particular, the choice adopted is the simplest
permitting to fulfill the Dirichlet boundary conditions [imposed on the
fluid fields]. This is obtained prescribing the functional form of $f_{i}(%
\mathbf{r,}t)$ on the boundary of the fluid domain ($\delta \Omega $), which
is identified with a function $f_{oi}(\mathbf{r},t).$

Second, the functional class of $f_{i}(\mathbf{r,}t),$ $f_{i}^{eq}(\mathbf{r,%
}t)$ and of $f_{oi}(\mathbf{r},t)$ remains essentially arbitrary. Thus, in
particular, the initial and boundary conditions, specified by the same
function $f_{oi}(\mathbf{r},t),$ can be defined imposing the positions (\ref%
{Eq.11d}),(\ref{Eq.11e}). As further basic consequence, $f_{i}^{eq}(\mathbf{%
r,}t)$ and $f_{i}(\mathbf{r,}t)$ need not necessarily be Galilei-invariant
(in particular they may not be invariant with respect to velocity
translations), although the fluid equations must be necessarily fully
Galilei-covariant. As a consequence it is always possible to select\ $%
f_{i}^{eq}(\mathbf{r,}t)$ and $f_{oi}(\mathbf{r},t)$ based on convenience
and mathematical simplicity. Thus, besides distributions which are Galilei
invariant and satisfy a principle of maximum entropy (see for example \cite%
{Karlin1998,Karlin1999,Ansumali2000,Ansumali2002,Boghosian2001,Ansumali2003}%
), it is always possible to identify them [i.e., $\ f_{i}^{eq}(\mathbf{r,}%
t),f_{oi}(\mathbf{r},t)$] with a non-Galilean invariant polynomial
distribution of the type (\ref{Polynomial}) [manifestly, to be exactly
Galilei-invariant each\ $f_{i}^{eq}(\mathbf{r,}t)$ should depend on velocity
only via the relative velocity $\mathbf{u}_{i}=\mathbf{a}_{i}-\mathbf{V}$].

We mention that the non-uniqueness of the source term $\widetilde{S}_{i}$
can be exploited also by imposing that $f_{i}^{eq}(\mathbf{r,}t)$ results a
particular solution of the inverse kinetic equation Eq.(\ref{Eq.7}) and
there results also $f_{oi}(\mathbf{r},t)=f_{i}^{eq}(\mathbf{r,}t)$. In
Appendix B we report the extension of THM.1 which is obtained by identifying
again $f_{i}^{eq}(\mathbf{r,}t)$ with the polynomial distribution (\ref%
{Polynomial}).

\section{6 - The entropic principle - Condition of positivity of the kinetic
distribution function}

A fundamental limitation of the standard LB approaches is their difficulty
to attain low viscosities, due to the appearance of numerical instabilities
\cite{Succi}. In numerical simulations based on customary LB approaches
large Reynolds numbers is usually achieved by increasing numerical accuracy,
in particular strongly reducing the time step and the grid size of the
spatial discretization (both of which can be realized by means of numerical
schemes with adaptive time-step and using grid refinements). Hence, the
control [and possible inhibition] of numerical instabilities is achieved at
the expense of computational efficiency. This obstacle is only partially
alleviated by approaches based on ELBM \cite%
{Karlin1998,Karlin1999,Ansumali2000,Ansumali2002,Boghosian2001,Ansumali2003}%
.\ Such methods are based on the hypothesis of fulfilling an H-theorem,
i.e., of satisfying in the whole domain $\Omega \times I$ the condition of
strict positivity for the discrete kinetic distribution functions$.$ \ This
requirement is considered, by several authors (see for example \cite%
{Succi2002,Boghosian2003,Chikatamarla2006}), an essential prerequisite to
achieve numerical stability in LB simulations. However, the numerical
implementation of ELBM typically induce a substantial complication of the
original algorithm, or require a cumbersome fine-tuning of adjustable
parameters \cite{Lallemand2000,Ansumali2002}.

\subsection{6A - The constant entropy principle and PEM}

A basic aspect of the IKT's here developed is the possibility of fulfilling
identically the strict positivity requirement by means of a suitable
H-theorem which provides also a maximum entropy principle. In particular, in
this Section,\ extending the results of THM.1 and 2, we intend to prove that
\emph{a constant H-theorem can be established both for the integral and
differential LB-IKT's defined above}. The H-theorem can be reached by
imposing for the Gibbs-Shannon entropy functional the requirement that for
all $t\in I$ there results
\begin{equation}
\frac{\partial }{\partial t}S(f)=-\frac{\partial }{\partial t}%
\int\limits_{\Omega }d^{3}r\sum\limits_{i=0,8}f_{i}\ln
(f_{i}/w_{i})=0, \label{Eq.21}
\end{equation}%
which implies that $S(f)$ is necessarily maximal in a suitable functional
set $\left\{ f\right\} .$ The result can be stated as follows:

\subsection{Theorem 2 - \emph{Constant H-theorem}}

\emph{In validity of THM.1, let us assume that:}

\emph{1) the configuration domain }$\Omega $ \emph{is bounded;}

\emph{2) at time }$t_{o}$\emph{\ the discrete kinetic distribution functions
}$f_{i},$ \emph{for }$i=0,8,$ \emph{are all strictly positive in the set} $%
\overline{\Omega }.$

\emph{Then the following statements hold:}

\emph{A) by suitable definition of the} \emph{pseudo pressure
}$P_{o}(t),$ \emph{the Gibbs-Shannon entropy functional
}$S(f)=-\int\limits_{\Omega }d^{3}r\sum\limits_{i=0,8}f_{i}\ln
(f_{i}/w_{i})$ \emph{can be set to be} \emph{constant in the whole
time interval }$I.$ \emph{This holds provided
the pseudo-pressure }$P_{o}(t)$\emph{\ satisfies the differential equation }%
\begin{eqnarray}
&&\left. \frac{\partial P_{o}}{\partial t}\int\limits_{\Omega
}d^{3}r\sum_{i=0}^{8}\frac{w_{i}}{c^{2}}\left( 1+\log f_{i}\right)
=\right.
\label{Eq.22} \\
&=&\int\limits_{\Omega }d^{3}r\sum_{i=0}^{8}\left(
\mathbf{a}_{i}\cdot \nabla f_{i}-\widehat{S}_{i}\right) \left(
1+\log f_{i}\right) ,  \notag
\end{eqnarray}%
\emph{where }$\widehat{S}_{i}=S_{i}+\frac{w_{i}}{c^{2}}\frac{\partial P_{o}}{%
\partial t}$;

\emph{B) if the entropy functional }$S(f)=-\int\limits_{\Omega
}d^{3}r\sum\limits_{i=0,8}f_{i}\ln (f_{i}/w_{i})$ \emph{is
constant in} \emph{the whole time interval }$I$ \emph{the discrete
kinetic distribution functions} $f_{i}$ \emph{are all strictly
positive in the whole set} $\Omega \times I;$

\emph{C) an arbitrary solution of LB-IKE [Eq.(\ref{Eq.7})] which satisfies
the requirement A) is extremal in a suitable functional class and maximizes
the Gibbs-Shannon entropy .}

\textbf{Proof:}

A) Invoking Eq.(\ref{Eq.7}), there results%
\begin{eqnarray}
&&\left. \frac{\partial S(t)}{\partial t}=-\int\limits_{\Omega
}d^{3}r\sum_{i=0}^{8}\frac{\partial f_{i}}{\partial t}\left[ 1+\log f_{i}%
\right] =\right.  \label{Eq.23} \\
&=&\int\limits_{\Omega }d^{3}r\sum_{i=0}^{8}\left(
\mathbf{a}_{i}\cdot \nabla f_{i}-S_{i}\right) \left( 1+\log
f_{i}\right) ,  \notag
\end{eqnarray}

where $S_{i}$ is the source term, provided by Eq.(\ref{Eq.14}).\ By direct
substitution it follows the thesis.

B) If Eq.(\ref{Eq.22}) holds identically in there results $\forall t\in
I,S\left( t\right) =S\left( t_{0}\right) ,$ which implies the strict
positivity of $f_{i},$ for all\emph{\ }$i=0,8.$

C) Let us introduce the functional class
\begin{equation}
\left\{ f+\alpha \delta f\right\} =\left\{ f_{i}=f_{i}(t)+\alpha \delta
f_{i}(t),i=0,8\right\} ,  \label{Eq.24}
\end{equation}%
where $\alpha $ is a finite real parameter and the synchronous variation $%
\delta f_{i}(t)$ is defined $\delta f_{i}(t)=df_{i}(t)\equiv \frac{\partial
f_{i}(t)}{\partial t}dt.$ Introducing the synchronous variation of the
entropy, defined by $\delta S\left( t\right) =\left. \frac{\partial }{%
\partial \alpha }\psi (\alpha )\right\vert _{\alpha =0},$ with $\psi (\alpha
)=S\left( f+\alpha \delta f\right) ,$ it follows%
\begin{equation}
\delta S\left( t\right) =dt\frac{\partial S(t)}{\partial t}.  \label{eq25}
\end{equation}

Since in validity of Eq.(\ref{Eq.22}) there results $\frac{\partial S(t)}{%
\partial t}=0,$ which in view of Eq.(\ref{eq25}) implies also $\delta
S\left( t\right) =0.$ It is immediately follows that there results
necessarily $\delta ^{2}S\left( t\right) \leq 0,$ i.e., $S\left( t\right) $
is maximal. Therefore, the kinetic distribution function which satisfies IKE
(Eq.(\ref{Eq.7})] is extremal in the functional class of variations (\ref%
{Eq.24}) and maximizes the Gibbs-Shannon entropy functional.

\subsection{6B - Implications}

In view of statement B, THM.2 warrants the strict positivity of the discrete
distribution functions $f_{i}$ ($i=0,8$) only in the open set $\Omega \times
I,$ while nothing can be said regarding their behavior on the boundary $%
\delta \Omega $ (on which $f_{i}$ might locally vanish)$.$ However, since
the inverse kinetic equation actually holds only in the open set $\Omega
\times I$, this does not affect the validity of the result. While the
precise cause of the numerical instability of LBM's is still unknown,the
strict positivity of the distribution function is usually considered
important for the stability of the numerical solution \cite%
{Boghosian2001,Succi2002}. It must be stressed that the numerical
implementation of the condition of constant entropy Eq.(\ref{Eq.22}) should
be straightforward, without involving a significant computational overhead
for LB simulations. Therefore it might represent a convenient scheme to be
adopted also for customary LB methods.

\section{7 - Asymptotic approximations and comparisons with previous CFD
methods}

A basic issue is the relationship with previous CFD numerical methods,
particularly asymptotic LBM's. Here we consider, for definiteness, only the
case of the integral LB-IKT introduced in Sec.5.\ Another motivation is the
possibility of constructing new improved asymptotic models, which satisfy
with prescribed accuracy the required fluid equations [INSE], of extending
the range of validity of traditional LBM's and fulfilling also the entropic
principle (see Sec.6). The analysis is useful in particular to establish on
rigorous grounds the consistency of previous LBM's.\ The connection [with
previous LBM's] can be reached by introducing appropriate asymptotic
approximations for the IKT's, obtained by assuming that suitable parameters
which characterize the IKT's are infinitesimal (or infinite) (\emph{%
asymptotic parameters}). A further interesting feature is the possibility of
constructing in principle a class of new asymptotic LBM's \emph{with
prescribed accuracy} , i.e., in which the distribution function (and the
corresponding momenta) can be determined with predetermined accuracy in
terms of perturbative expansions in the relevant asymptotic parameters.%
{\small \ }Besides recovering the traditional low-Mach number LBM's \cite%
{He1997,Abe1997,He2002}, which satisfy the isochoricity condition only in an
asymptotic sense and are closely related to the Chorin artificial
compressibility method, it is possible to obtain an improved asymptotic
LBM's which satisfy exactly the same equation.

We first notice that the present IKT is characterized by the\ arbitrary
positive parameters $\nu _{c},c$ and the initial value $P_{o}(t_{o}),$ which
enter respectively in the definition of the BGK operator [see (\ref{Eq.2a}%
)], the velocity momenta and equilibrium distribution function $f_{i}^{eq}$.
Both $c$ and $P_{o}(t_{o})$ must be assumed strictly positive, while, to
assure the validity of \ THM.2, $P_{o}(t_{o})$ must be defined so that (for
all $i=0,8$) $f_{i}^{eq}(\mathbf{r,}t_{o})>0$ in the closure $\overline{%
\Omega }.$ Thanks to THM.1.and 2 the new theory is manifestly valid for
arbitrary finite value of these parameters. \ This means that they hold also
assuming%
\begin{eqnarray}
&&\left. \nu _{c}\sim \frac{1}{o(\varepsilon ^{\alpha _{\nu }})},\right.
\label{26} \\
&&\left. c\sim \frac{1}{o(\varepsilon ^{\alpha _{c}})}\right. ,  \label{26aa}
\\
&&\left. P_{o}(t_{o})\sim o(\varepsilon ^{0}),\right.  \label{26a}
\end{eqnarray}%
where $\varepsilon $ denotes a strictly positive real infinitesimal, $\alpha
_{\nu },\alpha _{c}>0$ are real parameters to be defined, while the extended
fluid fields $\left\{ \rho ,\mathbf{V},p_{1}\right\} $ and the volume force $%
\mathbf{f}$ are all assumed independent of $\varepsilon .$ Hence,\ with
respect to $\varepsilon $ they scale
\begin{equation}
\rho _{o},\mathbf{V,}p_{1},\mathbf{f}\sim o(\varepsilon ^{0}).\   \label{26b}
\end{equation}%
As a result, for suitably smooth fluid fields (i.e., in validity of Axiom 1)
and appropriate initial conditions for $f_{i}(\mathbf{r},t)$, it is expected
that the first requirement actually implies in the whole set $\overline{%
\Omega }\times I$ the condition of closeness $f_{i}(\mathbf{r},t)\cong
f_{i}^{eq}(\mathbf{r},t)\left[ 1+o(\varepsilon )\right] ,$ consistent with
the LB Assumption \#4. To display meaningful comparisons with previous LBM's
let us introduce the further assumption that the fluid viscosity is small in
the sense%
\begin{equation}
\mu \sim o(\varepsilon ^{\alpha _{\mu }}),  \label{26c}
\end{equation}%
with $\alpha _{\mu }\geq 1$ another real parameter to be defined. Asymptotic
approximations for the corresponding LB-IKE [Eq.(\ref{Eq.7})] can be
directly recovered by introducing appropriate asymptotic orderings for the
contributions appearing in the source term $S_{i}=\widetilde{S}_{i}$. Direct
inspection shows that these are provided by the (dimensional) parameters
\begin{eqnarray}
&&\left. M_{p,a}^{eff}\equiv \frac{1}{c^{2}}\frac{\partial p}{\partial t}%
,\right.  \label{25aa} \\
&&\left. M_{p,b}^{eff}\equiv \frac{1}{c}\left\vert \nabla \cdot \underline{%
\underline{\mathbf{\Pi }}}-\nabla p\right\vert ,\right.  \label{25aab} \\
&&\left. M_{\mathbf{V}}^{eff}\equiv \frac{1}{c}\left\vert \mu \mathbf{\nabla
}^{2}\mathbf{V}\right\vert \right. .  \label{25aac}
\end{eqnarray}%
The first two $M_{p,a}^{eff}$ and $M_{p,b}^{eff}$ are here denoted
respectively as\emph{\ (first and second) pressure} \emph{effective Mach
numbers,} driven respectively by the pressure time-derivative and by the
divergence of the pressure anisotropy $\underline{\underline{\mathbf{\Pi }}}%
-p\underline{\underline{\mathbf{1}}}.$ Furthermore, $M_{\mathbf{V}}^{eff}$
is denoted as \emph{velocity effective Mach number}.\emph{\ } Physically
relevant examples [of asymptotic LBM's] can be achieved by introducing
suitable orderings in terms of the single infinitesimal $\varepsilon $ for
the parameters $M_{p,a}^{eff},M_{p,b}^{eff}$ and $M_{\mathbf{V}}^{eff}.$ We
stress that these orderings, in principle, can be introduced \emph{without }%
actually\emph{\ introducing restrictions on the fluid fields,} i.e.,
retaining the assumption that the extended fluid fields are independent of $%
\varepsilon .$ Interesting cases are provided by the asymptotic orderings
indicated below.

\subsection{7A - \ Small effective Mach numbers ($%
M_{p,a}^{eff},M_{p,b}^{eff} $ and $M_{\mathbf{V}}^{eff}$)}

An important aspect of LB theory is the possibility of constructing
asymptotic LBM's with prescribed accuracy with respect to the infinitesimal
parameter $\varepsilon $, in the sense that the fluid equations are
satisfied at least correct up to terms of order $o(\varepsilon ^{n})$
included, with $n=1$ or $2,$ namely ignoring error terms of order $%
o(\varepsilon ^{n+1})$ or higher. Let us, first, consider the case in which
all parameters $M_{p,a}^{eff},M_{p,b}^{eff}$ and $M_{\mathbf{V}}^{eff}$ are
all infinitesimal w.r. to $\varepsilon $ (\emph{low-effective-Mach numbers}%
). Since the parameters $c$ and $\nu _{c}$ are free, they can be defined so
that that there results $c\sim \nu _{c}\sim 1/o(\varepsilon )$ [which
implies $\alpha _{c}=\alpha _{\nu }=1$]. This requires%
\begin{equation}
M_{p,a}^{eff}\sim M_{p,b}^{eff}\sim o(\varepsilon ^{2}).  \label{ordering A}
\end{equation}%
If{\small , }we consider a low-viscosity fluid for which the kinematic
viscosity $\nu =\mu /\rho _{o}$ can be assumed of order $\varepsilon $ [and
hence $\alpha _{\mu }=1$] it follows that
\begin{equation}
M_{\mathbf{V}}^{eff}\sim o(\varepsilon ^{2}).  \label{ordering B}
\end{equation}%
Thanks to the assumptions (\ref{26})-(\ref{26c}) there follows $\nabla \cdot
\underline{\underline{\mathbf{\Pi }}}-\nabla p\sim o(\varepsilon )$ and $\mu
\mathbf{\nabla }^{2}\mathbf{V}\sim o(\varepsilon ),$which implies that the
source term $\widetilde{S}_{i},$ ignoring corrections of order $%
o(\varepsilon ^{2}),$ becomes%
\begin{eqnarray}
\widetilde{S}_{i} &\cong &\widetilde{S}_{Ai}\left[ 1+o(\varepsilon )\right] ,
\label{7A-1} \\
\widetilde{S}_{Ai} &\equiv &-\frac{w_{i}}{c^{2}}\mathbf{a}_{i}\cdot \mathbf{f%
}.
\end{eqnarray}%
It is immediate to determine the corresponding moment equations, which read:%
\begin{eqnarray}
&&\left. \frac{1}{c^{2}}\frac{\partial p_{1}}{\partial t}+\nabla \cdot
\mathbf{V}=0,\right.  \label{7A-2} \\
&&\left. N\mathbf{V}=\mathbf{0}+o(\varepsilon ^{2}),\right.  \label{7A-3}
\end{eqnarray}%
Formally the first equation can be interpreted as an evolution equation for
the kinetic pressure $p_{1}.$ Nevertheless, in view of the ordering (\ref%
{ordering A}) it actually implies the isochoricity condition%
\begin{equation}
\nabla \cdot \mathbf{V}=0+o(\varepsilon ^{2}).  \label{7A-4}
\end{equation}%
Instead, the second one [Eq.(\ref{7A-3})]. due to the asymptotic
approximation (\ref{ordering B}), reduces to the Euler equation. Therefore
in this case the asymptotic approximation (\ref{7A-1}) is not adequate. To
recover the correct Navier-Stokes equation a more accurate approximation is
needed, realized requiring that the hydrodynamic equations are satisfied
correct to order $o(\varepsilon ^{3}).$ A fist possibility is to consider a
more accurate approximation for the source term. Restoring the pressure and
viscous source terms in (\ref{7A-1}) there results the asymptotic source
term
\begin{equation}
\widetilde{S}_{Bi}\equiv \frac{w_{i}}{c^{2}}\left[ \frac{\partial p_{1}}{%
\partial t}-\mathbf{a}_{i}\cdot \left( \mathbf{f}_{1}\mathbf{-}\mu \mathbf{%
\nabla }^{2}\mathbf{V}\right) \right] ,  \label{7A-5}
\end{equation}%
where in validity of the previous orderings
\begin{equation}
\widetilde{S}_{i}\cong \widetilde{S}_{Bi}\left[ 1+o(\varepsilon )\right] .
\end{equation}%
The corresponding moment equations become therefore%
\begin{eqnarray}
&&\left. \nabla \cdot \mathbf{V}=0,\right.  \label{7A-5b} \\
&&\left. N\mathbf{V}=\mathbf{0}+o(\varepsilon ^{3}).\right.  \label{7A-5c}
\end{eqnarray}%
It is remarkable that in this case the isochoricity condition is exactly
fulfilled, even if the source term is not the exact one. For the sake of
reference, it is interesting to mention another possible small-Mach-number
ordering. This is obtained imposing for the parameters $c$ and $\nu _{c}$
\begin{eqnarray}
c &\sim &\frac{1}{o(\varepsilon )},  \label{7A-6} \\
\nu _{c} &\sim &\frac{1}{o(\varepsilon ^{2})},  \label{7A-7}
\end{eqnarray}%
while requiring for $\nu =\mu /\rho _{o}$ the same constraint adopted by
asymptotic LBM's, namely Eq.(\ref{2bb-b}). In this case one can show that
the moment equation (\ref{7A-5c}) is actually satisfied correct to order $%
o(\varepsilon ^{3}),$ \emph{while the isochoricity condition is only
satisfied to order} $o(\varepsilon ^{2})$. The following theorem can, in
fact, be proven:

\subsection{Theorem 3 - \emph{Low effective-Mach-numbers asymptotic
approximation}}

\emph{In validity of THM.1, let us invoke the following assumptions:}

\emph{1) LB assumptions \#3 and \#4 for the discrete kinetic distributions }$%
f_{i}$\emph{\ (}$i=0,8$\emph{);}

\emph{2) the free parameters }$c$\emph{\ and }$\nu _{c}$\emph{\ are assumed
to satisfy the asymptotic orderings (\ref{7A-6}),(\ref{7A-7});}

\emph{3) the fluid viscosity }$\mu $\emph{\ \ is assumed of order }$\mu \sim
o(\varepsilon )$

\emph{4) the fluid viscosity }$\mu $\emph{\ \ is prescribed so that the
kinematic viscosity }$\nu =\mu /\rho _{o}\ $\emph{is defined in accordance
to Eq.(\ref{2bb-b});}

\emph{5) the kinetic pressure }$p_{1}$\emph{\ is assumed slowly varying in
the sense}%
\begin{equation}
\frac{\partial \ln p_{1}}{\partial t}\sim o(\varepsilon ).  \label{7A-8}
\end{equation}

\emph{It follows that the source term is approximated by Eq.(\ref{7A-1}) and
moment equations are provided by the asymptotic equations:}%
\begin{eqnarray}
&&\left. \frac{1}{c^{2}}\frac{\partial p_{1}}{\partial t}+\nabla \cdot
\mathbf{V}=0+o(\varepsilon ^{3}),\right.  \label{7A-9} \\
&&\left. N\mathbf{V}=\mathbf{0}+o(\varepsilon ^{3}),\right.  \label{7A-10}
\end{eqnarray}%
\emph{i.e., the isochoricity and NS equation are recovered respectively
correct to order }$o(\varepsilon ^{2})$ \emph{and} $o(\varepsilon ^{3}).$%
\emph{\ }

\textbf{Proof}

First we notice that the ordering assumptions 2)-5) require%
\begin{eqnarray}
M_{p,a}^{eff} &\sim &o(\varepsilon ^{3}) \\
M_{\mathbf{V}}^{eff} &\sim &o(\varepsilon ^{2}), \\
M_{p,b}^{eff} &\sim &o(\varepsilon ^{4}),
\end{eqnarray}%
which imply at least the validity of Eqs.(\ref{7A-1})-(\ref{7A-3}). The
proof of Eqs.(\ref{7A-9}) and (\ref{7A-10}) is immediate. In both cases it
sufficient to notice that in validity of hypotheses 1)-3) and in terms of a
Chapman-Enskog perturbative solution of Eq.(\ref{Eq.7}) there results
actually

\begin{equation}
\mathbf{-}\mu \mathbf{\nabla }^{2}\mathbf{V}-\nabla \cdot \underline{%
\underline{\mathbf{\Pi }}}+\nabla p=O+o(\varepsilon ^{3}),
\end{equation}%
and hence $\widetilde{S}_{i}$ reduces to Eq.(\ref{7A-1}).

The predictions of THM.3 are relevant for comparisons and to provide
asymptotic accuracy estimates for previous asymptotic LBM's [see Refs. \cite%
{He1997,Abe1997,He2002}]. In fact, the asymptotic moment equations (\ref%
{7A-9}) and (\ref{7A-10}) formally coincide with the analogous moment
equations predicted by such theories, when the kinetic pressure $p$ is
replaced by the fluid pressure $p_{1}$ (i.e., if the function $P_{o}(t)$ is
set identically equal to zero). \cite{He1997,Abe1997,He2002}. Nevertheless,
the accuracy of customary LBM's depends on the properties of the solutions
of INSE. In fact, if one assumes
\begin{equation}
\frac{\partial \ln p_{1}}{\partial t}\sim o(\varepsilon ^{0})
\end{equation}
the customary $(\mathbf{V},p)$ asymptotic LBM \cite{He1997,Abe1997,He2002}
result\emph{\ actually accurate only to order }$o(\varepsilon ^{2})$.
Therefore, in such case to reach an accuracy of order $o(\varepsilon ^{3})$
the approximation (\ref{7A-5}) must be invoked for the source term.

The other interesting feature of Eqs.(\ref{7A-9}) and (\ref{7A-10}) is that
they provide a connection with the\ artificial compressibility method (ACM)
postulated by Chorin \cite{Chorin1967}, previously motivated merely on the
grounds of an asymptotic LBM \cite{He2002}. In fact, these coincides with
the Chorin's pressure relaxation equation where $c$ can be interpreted as
sound speed of the fluid.\ However - in a sense - this analogy is purely
formal and is only due to the neglect of the first pressure source term in $%
S_{i}$. It disappears altogether in Eq.(\ref{7A-5b}) if we adopt the more
accurate asymptotic source term (\ref{7A-5}).\ A further difference is
provided by the adoption of the kinetic pressure $p_{1}$\ which replaces the
fluid pressure $p$\ (used in Chorin approach). We stress that the choice of $%
p_{1}$\ here adopted$,$\ with $P_{o}(t)$\ determined by the entropic
principle, represents an important difference, since it permits to satisfy
everywhere in $\Omega \times I$\ the condition of strict positivity for the
discrete kinetic distribution functions.

\subsection{7B - Finite pressure-Mach number $M_{p,a}^{eff}$}

Another possible asymptotic ordering, usually not permitted by customary
asymptotic LBM's, is the one in which the test particle velocity is finite,
namely $c\sim o(\varepsilon ^{0}),$ the viscosity remains arbitrary and is
taken of order $\mu \sim o(\varepsilon ^{0})$ while again $\nu _{c}$ is
assumed $\nu _{c}\sim 1/o(\varepsilon ^{2})$ [i.e., $\alpha _{c}=\nu
_{c}=0,\alpha _{\nu }=2$]. \ In this case the pressure Mach $M_{p,a}^{eff}$
number results finite, while velocity and the second pressure Mach numbers
are considered infinitesimal, respectively of first and second order in $%
\varepsilon ,$ namely
\begin{eqnarray}
M_{p,a}^{eff} &\sim &o(\varepsilon ^{0}),  \notag \\
M_{\mathbf{V}}^{eff} &\sim &o(\varepsilon ),  \label{28a} \\
M_{p,b}^{eff} &\sim &o(\varepsilon ^{2}).  \notag
\end{eqnarray}%
\ To obtain the fluid equation with the prescribed accuracy, say of order $%
o(\varepsilon ^{2}),$ it is sufficient to approximate the source term $%
\widetilde{S}_{i}$ in terms of $\widetilde{S}_{i}\cong S_{Bi}^{(o)}\left[
1+o(\varepsilon ^{2})\right] .$ The set of asymptotic moment equations
coincide therefore with Eqs.(\ref{7A-5b}),(\ref{7A-5c}). Again, the
isochoricity condition is exactly fulfilled, while in this case the NS
equation is accurate only to order $o(\varepsilon ^{2}).$

\subsection{7C -\emph{\ }Small effective pressure-Mach numbers\emph{\ }($%
M_{p,a}^{eff},M_{p,b}^{eff}$) and finite velocity-Mach number ($M_{\mathbf{V}%
}^{eff}$)}

Finally, another interesting case is the one in which the fluid viscosity $%
\mu $ remains finite (strongly viscous fluid), i.e., in the sense $\mu \sim
o(\varepsilon ^{0})$ [i.e., $\alpha _{\mu }=0$] while both parameters $c$
and $\nu _{c}$ are suitably large, and respectively scale as $c\sim
1/o(\varepsilon ),$ $\nu _{c}\sim 1/o(\varepsilon ^{2})$ [i.e., $\alpha
_{c}=1,$ $\alpha _{\nu }=2$]. Due to assumptions (\ref{26})-(\ref{26c}) one
obtains $\nabla \cdot \underline{\underline{\mathbf{\Pi }}}-\nabla p\sim
o(\varepsilon ^{2})$ and $\mu \mathbf{\nabla }^{2}\mathbf{V}\sim
o(\varepsilon ^{0}).$ It follows that the effective Mach numbers scale
respectively as

\begin{eqnarray}
M_{p,b}^{eff} &\sim &o(\varepsilon ^{3})  \label{27a} \\
M_{p,a}^{eff} &\sim &M_{\mathbf{V}}^{eff}\sim o(\varepsilon ^{2}),  \notag
\end{eqnarray}%
If we impose on $\mu $ also the same constraint set by Eq.(\ref{2bb-b}), the
customary asymptotic LBM's can be invoked also in this case. However, since
the first pressure and velocity Mach numbers are only second order accurate,
the NS equation is recovered to order $o(\varepsilon ^{2})$ only. \
Nevertheless, it is possible to recover \emph{with prescribed accuracy} the
fluid equations (\ref{7A-5b}),(\ref{7A-5c}). This is obtained adopting the
source term $\widetilde{S}_{i}\cong \widetilde{S}_{Bi}$ [see Eq.(\ref{7A-5}%
)]. As a basic consequence, the isochoricity equation is satisfied exactly
(hence no meaningful analogy with Chorin's approach arises), while the NS
equation results correct to order $o(\varepsilon ^{3}).$ These results
provide a meaningful extension of the customary asymptotic LBM's. We stress
that the entropic approach here developed holds independently of the
asymptotic orderings here considered [for\ the parameters $%
M_{p,a}^{eff},M_{p,b}^{eff},M_{\mathbf{V}}^{eff}$]. Thus it can be used in
all cases to assure the strict positivity of the discrete distribution
function.

\section{8 - Conclusions}

In this paper we have presented the theoretical foundations of a new
phase-space model for incompressible isothermal fluids, based on a
generalization of customary lattice Boltzmann approaches.{\small \ }We have
shown that many of the limitations of traditional (asymptotic) LBM's can be
overcome. \ As a main result, we have proven that the \emph{LB-IKT}\textit{\
\ }can be developed in such a way that it furnishes exact Navier-Stokes and
Poisson solvers, i.e., it is - in a proper sense - an inverse kinetic theory
for INSE. \ The theory exhibits several features, in particular we have
proven that the integral LB-IKT (see Sec.5):

\begin{enumerate}
\item determines uniquely the fluid pressure $p(\mathbf{r},t)$ via the
discrete kinetic distribution function without solving explicitly (i.e.,
numerically) the Poisson equation for the fluid pressure. Although analogous
to traditional LBM's, this is interesting since it is achieved without
introducing compressibility and/or thermal effects. In particular the
present theory does not rely on a state equation for the fluid pressure.

\item is \emph{complete}, namely all fluid fields are expressed as momenta
of the distribution function and all hydrodynamic equations are identified
with suitable moment equations of the LB inverse kinetic equation.

\item allows arbitrary initial and boundary conditions for the fluid fields.

\item is \emph{self-consistent}: the kinetic theory holds for arbitrary,
suitably smooth initial conditions for the kinetic distribution function. In
other words, the initial kinetic distribution function must remain arbitrary
even if a suitable set of its momenta are prescribed at the initial time.

\item the associated the kinetic and equilibrium distribution functions can
always be chosen to belong to the class of non-Galilei-invariant
distributions. In particular the equilibrium kinetic distribution can always
be identified with a polynomial of second degree in the velocity.

\item is \emph{non-asymptotic}, i.e., unlike traditional LBM's it does not
depend on any small parameter, in particular \emph{it holds for finite Mach
numbers}.

\item fulfills an entropic principle, based on a constant-H theorem. This
theorem assures, at the same time, the \emph{strict positivity of the
discrete kinetic distribution function} and the maximization of the
associated Gibbs-Shannon entropy in a properly defined functional class.
Remarkably the constant H-theorem is fulfilled for \emph{arbitrary (strictly
positive) kinetic equilibria}. This includes also the case of polynomial
kinetic equilibria.
\end{enumerate}

A further remarkable aspect of the theory concerns the choice of
the kinetic boundary conditions to be satisfied by the
distribution function (Axiom II) and obtained by prescribing the
form of the incoming-velocity distribution [see
Eq.(\ref{Eq.11a})]. Thanks to Eqs.(\ref{Eq.10b}),(\ref{Eq.10c}),
this requirement {[of the LB-IKT]} the boundary conditions for the
fluid fields are satisfied exactly while the fluid equations are
by construction identically fulfilled also arbitrarily close to
the boundary. This result, in a proper sense, applies only to
Dirichlet boundary conditions for the fluid fields [see
Eqs.(\ref{INSE-R2})]. Nevertheless the same approach can be in
principle extended to the case of mixed or Neumann boundary
conditions for the fluid fields.

Moreover, we have shown that a useful implication of the theory is provided
by the possibility of constructing asymptotic approximations to the inverse
kinetic equation. This permits to develop a new class of asymptotic LBM's
which satisfy INSE \emph{with prescribed accuracy,} to obtain useful
comparisons with previous CFD methods (Chorin's ACM) and to achieve accuracy
estimates for customary asymptotic LBM's. The main results of the paper are
represented by THM's 1-3, which refer respectively to the construction of
the integral LB-IKT, to the entropic principle and to construction of the
low effective-Mach-numbers asymptotic approximations. For the sake of
reference, also another type of LB-IKT, which admits as exact particular
solution the polynomial kinetic equilibrium, has been pointed out (THM.1bis).

The construction of a discrete inverse kinetic theory of this type for the
incompressible Navier-Stokes equations represents an exciting development
for the phase-space description of fluid dynamics, providing a new starting
point for theoretical and numerical investigations based on LB theory. In
our view, the route to more accurate, higher-order LBM's, here pointed out,
will be important in order to achieve substantial improvements in the
efficiency of LBM's in the near future.

\section{Appendix A}

The basic argument regarding the accuracy of the boundary conditions adopted
by customary asymptotic LBM's is provided by Ref.\cite{Skordos1993}. In
fact. let us assume that on the boundary $\delta \Omega $ the incoming
distribution function $f_{i}^{(-)}(\mathbf{r}_{w},t)$ is prescribed
according to Eqs.(\ref{Eq.10a}),(\ref{Eq.11b}) and (\ref{Eq.11c}), being $%
f_{oi}^{(-)}(\mathbf{r}_{w},t)$ prescribed suitably smooth functions which
are non vanishing only only for incoming discrete velocities $\mathbf{a}_{i}$
for which $\left( \mathbf{a}_{i}-\mathbf{V}_{w}\right) \cdot \mathbf{n}(%
\mathbf{r}_{w},t)\leq 0$. For definiteness, let us assume that $f_{oi}^{(-)}(%
\mathbf{r}_{w},t)\equiv f_{i}^{eq}(\mathbf{r}_{w},t)$ where $f_{i}^{eq}(%
\mathbf{r}_{w},t)$ denotes a suitable equilibrium distribution. It follows
that suitably close to the boundary the kinetic distribution differs from
the Chapman-Enskog solution (\ref{Eq.2}). The numerical error can be
overcome only discarding the first few spatial grid (close to the boundary)
in the numerical simulation \cite{Skordos1993}.

\section{Appendix B}

Unlike standard kinetic theory, the distinctive feature of LB-IKT's is the
possibility of adopting a non-Galilei invariant kinetic distribution
function (i.e., non-invariant with respect to velocity translations). \ Here
we report another example of discrete inverse kinetic theory of this type.
Let us modify Axiom IV so that to\ permit that a particular solution of
LB-IKE [Eq.(\ref{Eq.7})] is provided by $f_{i}=f_{i}^{eq}.$ Here we identify
$f_{i}^{eq}$ with the (non-Galilei invariant) polynomial kinetic
distribution defined by Eq.(\ref{Polynomial})\ but with the kinetic pressure
$p_{1}$\ that replace the fluid pressure $p$. In this case one can prove
that the source term $S_{i}$ reads
\begin{equation}
S_{i}=S_{i}^{(1)}\equiv \widetilde{S}_{i}+\Delta S_{i},  \label{Eq.15}
\end{equation}%
where%
\begin{eqnarray}
&&\left. \Delta S_{i}=\frac{w_{i}\rho _{o}}{c^{2}}\left[ \left( \mathbf{a}%
_{i}-\mathbf{V}\right) \cdot \nabla \mathbf{V-}\frac{1}{2}\mathbf{a}%
_{i}\nabla \mathbf{\cdot V}\right] \cdot \mathbf{a}_{i}+\right.  \notag \\
&&+\frac{w_{i}}{c^{2}}N_{1}\mathbf{V\cdot }\left[ \mathbf{V}-\mathbf{a}_{i}%
\frac{3\mathbf{a}_{i}\cdot \mathbf{V}}{c^{2}}\right] +  \label{Eq.16} \\
&&+\frac{1}{2}w_{i}\rho _{o}\mathbf{a}_{i}\cdot \nabla \left[ 3\left( \frac{%
\mathbf{a}_{i}\cdot \mathbf{V}}{c^{2}}\right) ^{2}-\frac{V^{2}}{c^{2}}\right]
.  \notag
\end{eqnarray}%
Here $N_{1}\equiv N-\rho _{o}\frac{\partial }{\partial t},$ where $N$ is the
Navier-Stokes operator (\ref{Eq.0}), namely $N_{1}$ is the nonlinear
operator which acting on $\mathbf{V}$ yields $N_{1}\mathbf{V}=\rho _{o}%
\mathbf{V}\cdot \nabla \mathbf{V+}\nabla \left[ p_{1}-\Phi \left( \mathbf{r}%
\right) \right] +\mathbf{f}_{1}\mathbf{-}\mu \mathbf{\nabla }^{2}\mathbf{V}.$
Hence, invoking INSE, $\Delta S_{i}$ can also be written in the equivalent
form%
\begin{eqnarray}
&&\left. \Delta S_{i}=\frac{w_{i}\rho _{o}}{c^{2}}\left[ \left( \mathbf{a}%
_{i}-\mathbf{V}\right) \cdot \nabla \mathbf{V-}\frac{1}{2}\mathbf{a}%
_{i}\nabla \mathbf{\cdot V}\right] \cdot \mathbf{a}_{i}+\right.  \notag \\
&&+\frac{w_{i}}{c^{2}}\rho _{o}\frac{\partial }{\partial t}\mathbf{V\cdot }%
\left[ \mathbf{V}-\mathbf{a}_{i}\frac{3\mathbf{a}_{i}\cdot \mathbf{V}}{c^{2}}%
\right] +  \label{Eq.17} \\
&&+\frac{1}{2}w_{i}\rho _{o}\mathbf{a}_{i}\cdot \nabla \left[ 3\left( \frac{%
\mathbf{a}_{i}\cdot \mathbf{V}}{c^{2}}\right) ^{2}-\frac{V^{2}}{c^{2}}\right]
.  \notag
\end{eqnarray}%
The following result holds:

\subsection{Theorem 1bis - \emph{Differential LB-IKT}}

\emph{In validity of axioms I-IV and the assumption that }$f_{i}=f_{i}^{eq}$%
\emph{\ is a particular solution of Eq.(\ref{Eq.7}),} \emph{the following
statements hold:}

\emph{A) }$\ f_{i}^{eq}$\emph{\ \ is a particular solution of LB-IKE [Eq.(%
\ref{Eq.7})] if and only if the extended fluid fields }$\left\{ \mathbf{V,}%
p_{1}\right\} $\emph{\ are strong solutions of INSE of class (\ref{Eq.6a}),
with initial and boundary conditions (\ref{INSE-R1})-(\ref{INSE-R2}), and
arbitrary pseudo pressure }$p_{o}(t)$\emph{\ of class }$C^{(1)}(I).$\emph{\ }

\emph{Moreover, for an arbitrary particular solution }$f_{i}$\emph{\ and for
arbitrary extended fluid fields}$:$

\emph{For an arbitrary particular solution }$f_{i}:$\emph{\ }

\emph{B) }$\ f_{i}$\emph{\ \ is a solution of LB-IKE [Eq.(\ref{Eq.7})] if
and only if the extended fluid fields }$\left\{ \mathbf{V,}p_{1}\right\} $%
\emph{\ are arbitrary strong solutions of INSE of class (\ref{Eq.6a}), with
initial and boundary conditions (\ref{INSE-R1})-(\ref{INSE-R2}), and
arbitrary pseudo pressure }$p_{o}(t)$\emph{\ of class }$C^{(1)}(I);$

\emph{C) the moment equations of L-B IKE coincide identically with INSE in
the set }$\Omega \times I;$

\emph{D) the initial conditions and the (Dirichlet) boundary conditions for
the fluid fields are satisfied identically;}

\emph{E) the source term }$S_{i}$\emph{\ is uniquely defined by Eqs.(\ref%
{Eq.15}),(\ref{Eq.16});}

\textbf{Proof:}

The proof of propositions A,B, C and D is analogous to that provided in
THM.1. Assuming $S_{i}=S_{i}^{(1)},$ the proof of B follows from
straightforward algebra. In fact, letting $f_{i}(\mathbf{r},t)=f_{i}^{eq}(%
\mathbf{r},t)$ for all $(\mathbf{r},t)\in \overline{\Omega }\times I$ in the
LB-IKE [Eq.(\ref{Eq.7})], one finds that Eq.(\ref{Eq.7}) is fulfilled iff
the fluid fields satisfy the Navier-Stokes, isochoricity and
incompressibility equations (\ref{INSE-1}),(\ref{INSE-2}) and (\ref{INSE-3}%
). The proof of proposition E can be reached in a similar way. The
uniqueness of the source term $S_{i}$ is an immediate consequence of the
uniqueness of the solutions for INSE.

\bigskip

\textbf{ACKNOWLEDGEMENTS} %\begin{acknowledgement}
Useful comments and stimulating discussions with K.R. Sreenivasan, Director,
ICTP (International Center of Theoretical Physics, Trieste, Italy) are
warmly acknowledged. Research developed in the framework of PRIN Project
\textit{Fundamentals of kinetic theory and applications to fluid dynamics,
magnetofluid dynamics and quantum mechanics} (MIUR, Ministry for University
and Research, Italy), with the support of the Consortium for Magnetofluid
Dynamics, Trieste, Italy.

\end{document}